\documentclass[aps,floats,twocolumn,pre,10pt,groupedaddress,superscriptaddress]{revtex4-1}
\usepackage{amsmath,amssymb,amsthm,amsfonts,bm,bbm,braket,enumitem}
\usepackage[english]{babel}

\usepackage{natbib}
\usepackage{hyperref}
\hypersetup{
	pdfauthor={Cosimo Lupo and Federico Ricci-Tersenghi},
	pdftitle={Approximating the XY model on a random graph with a q-state clock model},
	colorlinks,
	linktocpage=true,
	pdfstartpage=1,
	pdfstartview=FitV,
	breaklinks=true,
    pageanchor=true,
    pdfpagemode=UseOutlines,
    plainpages=false,
    bookmarksnumbered,
    bookmarksopen=true,
    bookmarksopenlevel=1,
    hypertexnames=true,
    pdfhighlight=/O,
    urlcolor=blue,
    linkcolor=blue,
    citecolor=blue,
}

\allowdisplaybreaks

\usepackage[utf8x]{inputenc}	
\usepackage[T1]{fontenc} 	

\newcommand{\vs}{\vec{\sigma}}

\newcommand{\mD}{\mathcal{D}}
\newcommand{\mF}{\mathcal{F}}
\newcommand{\mG}{\mathcal{G}}
\newcommand{\mH}{\mathcal{H}}
\newcommand{\mM}{\mathcal{M}}
\newcommand{\mN}{\mathcal{N}}
\newcommand{\mP}{\mathcal{P}}

\newcommand{\mZ}{\mathcal{Z}}

\usepackage{graphicx}
\usepackage{tikz,tkz-berge,tkz-graph}
\usetikzlibrary{arrows,petri,topaths,calc}
\usepackage{array,booktabs}

\begin{document}

\title{Approximating the XY model on a random graph with a $q$-state clock model}

\author{Cosimo Lupo}
\affiliation{Dipartimento di Fisica, Sapienza Universit\`a di Roma, P.le A. Moro 5, I-00185 Rome, Italy}
\author{Federico Ricci-Tersenghi}
\affiliation{Dipartimento di Fisica, INFN-Sezione di Roma1, CNR-Nanotec Unit\`a di Roma, Sapienza Universit\`a di Roma, P.le A. Moro 5, I-00185 Rome, Italy}
\date{\today}

\begin{abstract}
Numerical simulations of spin glass models with continuous variables set the problem of a reliable but efficient discretization of such variables. In particular, the main question is how fast physical observables computed in the discretized model converge toward the ones of the continuous model when the number of states of the discretized model increases. We answer this question for the XY model and its discretization, the $q$-state clock model, in the mean-field setting provided by random graphs. It is found that the convergence of physical observables is exponentially fast in the number~$q$ of states of the clock model, so allowing a very reliable approximation of the XY model by using a rather small number of states. Furthermore, such an exponential convergence is found to be independent from the disorder distribution used. 
Only at $T=0$ the convergence is slightly slower (stretched exponential).

Thanks to the analytical solution to the $q$-state clock model, we compute accurate phase diagrams in the temperature versus disorder strength plane. We find that, at zero temperature, spontaneous replica symmetry breaking takes place for any amount of disorder, even an infinitesimal one.
We also study the one step of replica symmetry breaking (1RSB) solution in the low-temperature spin glass phase. 
\end{abstract}

\maketitle

\section{Introduction}

The theory of spin glasses, and disordered systems in general, has been mostly developed using the Ising variables~\cite{Book_MezardEtAl1987}. And the same is true also in different but related fields of research, such as discrete combinatorial optimization, where the Boolean variables play a prominent role~\cite{Book_MezardMontanari2009}.
The reason for this is not surprising: even with simple dichotomous variables, these models are very complex and show a very rich behavior. Thus the interest in moving to $m$-component variables living in a higher-dimensional space (e.\,g. the XY or Heisenberg spins), which would make computations harder, has been limited and this is witnessed by the scarcer literature available with respect to the Ising case.

Nonetheless, it is known from the experiments performed on spin glass materials having different degrees of anisotropy that the behavior of Ising-like models is different from Heisenberg-like models~\cite{BertEtAl2004}. This difference is particularly evident in some properties, like rejuvenation and memory effects, that are considered as a trademark of the complex and hierarchical organization of spin glass long-range order. Numerical simulations of Ising spin glasses at present have not found any clear evidence for rejuvenation~\cite{PiccoEtAl2001, MaioranoEtAl2005} and new extensive numerical simulations of XY or Heisenberg spin glasses will be required in the near future.

Up to now, most of the numerical simulations of XY or Heisenberg spin glasses have been performed on three-dimensional lattices with the main aim of understanding the role of the chiral long-range order~\cite{Villain1977, OliveEtAl1986, KawamuraTanemura1987, Kawamura1992, HukushimaKawamura2000, KawamuraLi2001, LeeYoung2003, CamposEtAl2006, FernandezEtAl2009}, but $m$-component variables naturally appear in other interesting problems, like the synchronization problems~\cite{Book_Kuramoto1975, AcebronEtAl2005, BandeiraEtAl2014, JavanmardEtAl2016} and the very recent field of random lasers~\cite{GordonFischer2002, AngelaniEtAl2006, AntenucciEtAl2015}, just for citing a few.

From the analytical point of view, models with $m$-component variables have been mostly studied on fully connected topologies~\cite{KirkpatrickSherrington1978, deAlmeidaEtAl1978, ElderfieldSherrington1982a, ElderfieldSherrington1982b}, often giving results quite different with respect to the ones of the Ising case. For example, the Gabay-Toulose critical line~\cite{GabayToulouse1981} exists only for $m>1$. A limiting case that has been studied in some detail is the one where the number of components diverges ($m\to\infty$). In this limit, some analytical computations can be performed \cite{AspelmeierMoore2004,beyer2012one,JavanmardEtAl2016}, although it is worth stressing that in the $m\to\infty$ limit the free energy landscape becomes much less complex (since the energy function is convex), and thus the phase diagrams simplify.  Also, the way the low-temperature physics changes by increasing the number of components is a very interesting problem that deserves specific studies \cite{baity2015inherent,ricci2016performance}. Here, we are interested in models with continuous variables, but with a small number of components: this is the reason why we choose to study the XY model ($m=2$).

A key difference between Ising ($m=1$) models and vector ($m>1$) models is that variables in the latter are continuous. This may be bothersome both in analytical and numerical computations: in the former case dealing with probability distributions on an $m$-dimensional unit sphere can not be done exactly and requires strong approximations~\cite{JavanmardEtAl2016}, while in numerical simulations working with discrete variables often allows one to better optimize the simulation code (e.\,g. by using look-up tables).
It is thus natural to ask how good can be a discrete approximation to a vector model.

It is well known that the discretization of a ferromagnetic system of vector spins in the low-temperature region works very badly, e.\,g. as it happens for SU(3) symmetry in lattice gauge theories~\cite{GrosseKuhnelt1982}. The main reason for this failure is the fact small thermal fluctuations around the fully ordered ground state are not well described by the discretized model.
However, when quenched disorder is introduced in the model, the situation may dramatically change. Indeed, in this case the presence of frustration makes low-energy configurations not fully ordered and  much more abundant: the inability of the discretized model to correctly describe small fluctuations may be not so relevant as long as it can cope with the many low-energy configurations.

To understand how good a discretized model can reproduce the physics of a vector model, in the present work we study the $q$-state clock model, which is a discretized version of the XY model.
We will mainly consider how physical observables change when increasing the number~$q$ of states and check how fast the XY model is approached in the $q\to\infty$ limit.
A similar question has been answered in a very recent work~\cite{MarruzzoLeuzzi2015} for the XY model with four-spin interactions, while here we only consider models with two-spin interactions, that belong to a different universality class with respect to those studied in Ref.~\cite{MarruzzoLeuzzi2015}.

More precisely, we are going to consider the $q$-state clock model in both the ferromagnetic and the spin glass versions with different kinds of disorder. We will focus on models defined on random regular graphs.
We will study phase diagrams in the temperature versus disorder strength plane at many values of $q$. Most of the computations are analytic, within the replica symmetric~(RS) ansatz. Finally, by considering the ansatz with one step of replica symmetry breaking~(1RSB), we will try to understand whether the universality class changes by varying~$q$.

A short comment on chiral ordering, which is a very debated issue in models with continuous variables defined on regular lattices \cite{kawamura1991chiral,LeeYoung2003,obuchi2013monte,baity2014phase}. When a model with continuous spin variables is defined on a random graph the topological defects play no longer any role. Indeed, typical loops in random graphs are $O(\ln{N})$ long, while short loops are rare, with a $O(1/N)$ density. Chiral ordering can not take place on random graphs, so we do not enter at all into the debate on the coupling/decoupling between spin and chiral degrees of freedom.

The structure of this paper is the following. In Sec.~\ref{sec:the_models} we introduce the models we are going to study, the XY model and its discretized version, the $q$-state clock model. Then in Sec.~\ref{sec:bp_cavity_method_rs}, we recall the basic features of cavity method in the replica symmetric framework and discuss about its validity. In Sec.~\ref{sec:rs_solution_xy_clock}, we solve the XY model and the $q$-state clock model on random regular graphs within the replica symmetric ansatz, by exploiting the cavity method both analytically and numerically. Then, in Sec.~\ref{sec:convergence}, we actually study the convergence of physical observables for the $q$-state clock model when~$q$ is increased, reaching the limiting values given by the XY model. In the end, in Sec.~\ref{sec:bp_cavity_method_1rsb}, we extend the cavity method to one step of replica symmetry breaking and we apply it to the $q$-state clock model, in order to give an insight of the exact solution (which should be full replica symmetry breaking) and to see if and when the universality class changes when $q$ increases.

\section{The models}
\label{sec:the_models}

\subsection{XY model}

The simplest case of continuous variables models --- or vector spin models --- is the \emph{XY model} whose Hamiltonian can be written either as
\begin{equation}
	\mathcal{H}[\{\vs\}] = -\sum_{\braket{ij}}J_{ij}\,\vs_i\cdot\vs_j\,,
\end{equation}
where spins have $m=2$ components ($\vs_i\in\mathbb{R}^2$) and unit norm ($|\vs_i|=1$), either as
\begin{equation}
	\mathcal{H}[\{\theta\}] = -\sum_{\braket{ij}}J_{ij} \cos(\theta_i-\theta_j)\,,
	\label{eq:H_xy_pm_J}
\end{equation}
with $\theta_i\in[0,2\pi)$.
The sums run over the pairs of nearest-neighbor vertices on a generic graph.
Couplings $J_{ij}$ can be all positive in ferromagnetic models or can be extracted from a distribution $\mathbb{P}_J(J_{ij})$ having support on both positive and negative regions in the spin glass case.

Despite of its simplicity, the XY model shows a lot of interesting features and allows new phenomena to rise up with respect to the Ising case: for example, thanks to the continuous nature of its variables, at very low temperatures small fluctuations are allowed in the XY model. In turn, this may produce null or very small eigenvalues in the spectral density of the Hessian matrix of the model. Other interesting features of the XY model --- and of vector spin models in general --- regard its behavior in an external field~\footnote{C.~Lupo, G.~Parisi, and F.~Ricci-Tersenghi, \textit{work in progress}.}, where e.\,g. one can observe different kinds of transitions: the de~Almeida-Thouless transition~\cite{deAlmeidaThouless1978} --- which is present also in the Ising case --- and the Gabay-Toulouse transition~\cite{GabayToulouse1981, CraggEtAl1982}, which does not show up in Ising models.

\subsection{$q$-state clock model}

The problem of discretizing the continuous variables of the XY model may naturally arise when one wishes to simulate very efficiently the model (e.\,g. by the use of look-up tables) or even when a reduction in the variables domain is required in the search for an analytically treatable solution~\cite{JavanmardEtAl2016}.
The simplest way to discretize the XY model is to allow each spin $\vs_i$ to take only a finite number~$q$ of directions along the unit circle, equally spaced by the elementary angle~$2\pi/q$. Taking $q$ large enough the error committed by the discretization should be negligible.

The Hamiltonian of the $q$-state clock model has formally the same expression in~(\ref{eq:H_xy_pm_J}), but with angles taking value in a finite set $\theta_i \in \{0,2\pi/q,4\pi/q,\ldots,2\pi(q-1)/q\}$.

While the XY model is recovered in the $q\to\infty$ limit, very small values of $q$ are expected to produce a rather different behavior: in particular, $q=2$ corresponds to the Ising model, the $q=3$ clock model can be mapped to a $3$-state Potts model and the $4$-state clock model is nothing but a \emph{double} Ising model, apart from a rescaling of couplings $J_{ij}$ by a factor $1/2$~\cite{NobreSherrington1986}.

For larger $q$ values, the XY model is approached and we will try to understand how fast is this process in the different parts of the phase diagram.
In particular we are interested in the role played by the disorder. Indeed, the slowest convergence is expected at very low temperatures, and in this region the different kinds of long range order (ferromagnetic or spin glass) may vary sensibly the convergence to the XY model.
Our naive expectation is that the strong frustration present in a spin glass phase may produce low-energy configurations in the clock model which are less rigid (with respect to the ferromagnetic case) and thus may have small energy fluctuations, so making the disordered $q$-state clock model more similar to the disordered XY model, with respect to the corresponding ferromagnetic versions.

\subsection{Random graphs}

Mean-field approximations are correct for models defined on fully connected graphs and on sparse random graphs; the latter case, apart from being more general (the fully connected topology can be recovered in the limit of large mean degree), is much more interesting, because the variables have a number of neighbors $O(1)$ and this produces fluctuations in the local environment of each variable, that closely resembles what happens in systems defined in a finite dimensional space.

In this work for simplicity we focus on models defined on random $c$-regular graphs (RRG), where each vertex has exactly $c$ neighbors. Given that we are going to solve the cavity equations via the population dynamics algorithm, we do not need to specify the algorithm to generate any specific realization of the RRG (e.\,g. the configurational model).

\section{Replica symmetric cavity method}
\label{sec:bp_cavity_method_rs}

Exact solutions of spin glass models on random graphs is highly non-trivial, due to the spatial fluctuations naturally induced by the topology of such graphs. Furthermore, as in the Sherrington-Kirkpatrick model, we expect the XY model to require a full replica symmetry breaking~(fRSB) scheme to be exactly solved~\cite{Book_MezardEtAl1987}. Anyway, a large class of models defined on random graphs can be solved just by exploiting a single step of replica symmetry breaking~(1RSB), as, e.\,g. the diluted $p$-spin~\cite{MezardEtAl2003}, the coloring problem, and the Potts model~\cite{ZdeborovaKrzakala2007, KrzakalaZdeborova2008}. In some cases, it can be demonstrated that such solutions are stable towards further steps of replica symmetry breaking~\cite{MontanariEtAl2004}.

Here we start from the simplest replica symmetric~(RS) ansatz, which actually corresponds to the well known Bethe approximation~\cite{Bethe1935}.
The RS ansatz is always correct for models defined on a tree and for models defined on a random graph if model correlations decay fast enough~\cite{Book_MezardMontanari2009}.
The locally tree-like structure of random graphs allows one to use the cavity method, either at the RS level~\cite{MezardParisi1987} or at the 1RSB level~\cite{MezardParisi2001, MezardParisi2003}.

The belief propagation~(BP) algorithm~\cite{Book_Pearl1988, YedidiaEtAl2003} is a convenient recursive algorithm to solve the RS self-consistency equations on a given graph. However, here we are interested in understanding the physical properties of typical random graphs, and this can be better achieved by the use of the population dynamics algorithm~\cite{Book_MezardMontanari2009}.

\subsection{Cavity equations at finite temperature}

The basic idea of cavity method is that on a tree the local marginals
\[
	\eta_i(\theta_i) = \sum_{\underline\theta \setminus \theta_i} \mu(\underline\theta)
	\qquad , \quad
	\mu(\underline\theta) = \frac1Z e^{-\beta \mH(\underline\theta)}
\]
can be written in term of cavity marginals $\eta_{i\to j}(\theta_i)$, namely the marginal probability of variable $\theta_i$ in the graph where the coupling $J_{ij}$ has been removed (hence the name cavity),
\begin{equation}
	\eta_{i}(\theta_i)=\frac{1}{\mZ_i}\,\prod_{k\in\partial i}\int d\theta_k\,e^{\,\beta J_{ik}\cos{(\theta_i-\theta_k)}}\,\eta_{k\to i}(\theta_k)\,,
	\label{eq:fullMarginals}
\end{equation}
where $\mZ_i$ is a normalization constant
\begin{equation}
	\mZ_i=\int d\theta_i\prod_{k\in\partial i}\int d\theta_k\,e^{\,\beta J_{ik}\cos{(\theta_i-\theta_k)}}\,\eta_{k\to i}(\theta_k)
\end{equation}
and $\partial i$ is the set of neighbors of spin $i$.

\begin{figure}[!t]
	\centering
	\includegraphics[width=0.85\columnwidth]{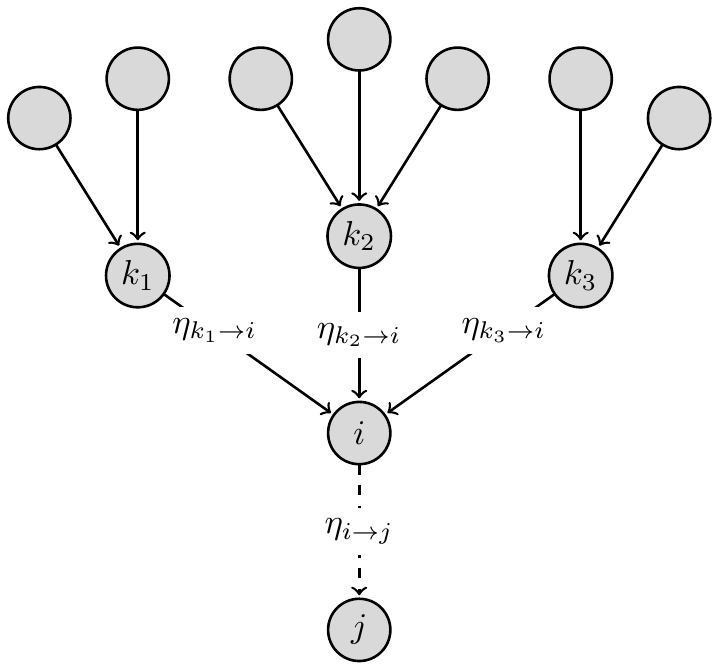}
	\caption{Due to the \emph{Bethe approximation}, when all edges around site $i$ are removed from the graph, then its neighbors become uncorrelated. So marginal probability distribution $\eta_{i\to j}(\theta_i)$ of site $i$ when edge $\braket{ij}$ has been removed from the graph can be computed iteratively from marginals $\eta_{k\to i}(\theta_k)$'s, $k\in\partial i \setminus j$. Alternatively, the \emph{belief} coming out from site $i$ to site $j$ is given by all other beliefs entering site $i$ from sites $k$'s belonging to $\partial i\setminus j$.}
	\label{fig:random_graphs}
\end{figure}

Referring to the notation in Fig.~\ref{fig:random_graphs}, it is not difficult to write down self-consistency equations among the cavity marginals \cite{YedidiaEtAl2003, Book_MezardMontanari2009}:
\begin{equation}
\begin{split}
	\eta_{i\to j}(\theta_i) &= \mF(\{\eta_{k\to i},J_{ik}\})\\
	&= \frac{1}{\mZ_{i\to j}}\,\prod_{k\in\partial i\setminus j}\int d\theta_k\,\eta_{k\to i}(\theta_k)\,e^{\,\beta J_{ik}\cos{(\theta_i-\theta_k)}}\,,
	\label{eq:self_cons_eta}
\end{split}
\end{equation}
where $\mZ_{i\to j}$ is just the normalization constant
\begin{equation}
	\mZ_{i\to j}=\int d\theta_i\prod_{k\in\partial i\setminus j}\int d\theta_k\,\eta_{k\to i}(\theta_k)\,e^{\,\beta J_{ik}\cos{(\theta_i-\theta_k)}}\,.
\end{equation}
We call the above BP equations, where BP stands both for Bethe-Peierls and Belief Propagation.

The addition of site $i$ and couplings $J_{ik}$ with $k\in\partial i \setminus j$, causes a free energy shift $\Delta F_{i\to j}$ in the system, which is directly related to the normalization constant $\mZ_{i\to j}$ via
\begin{equation}
	\mZ_{i\to j}=e^{\,-\beta\,\Delta F_{i\to j}}\,.
\end{equation}

Once the cavity marginals satisfying Eq.~(\ref{eq:self_cons_eta}) have been computed, the free energy density in a graph of $N$ vertices is given by
\begin{equation}
f_N(\beta)=-\frac{1}{\beta N}\,\Bigl(\sum_i\ln{\mZ_i}-\sum_{\braket{ij}}\ln{\mZ_{ij}}\Bigr)\,,
	\label{eq:f_rs_bp}
\end{equation}
where $\mZ_{ij}$ is given by
\begin{equation}
	\mZ_{ij}=\int d\theta_i d\theta_j\,e^{\,\beta J_{ij}\cos{(\theta_i-\theta_j)}}\,\eta_{i\to j}(\theta_i)\,\eta_{j\to i}(\theta_j)\,.
\end{equation}
As usual, the true free energy density $f(\beta)$ is obtained in the thermodynamic limit, $f(\beta)=\lim_{N\to\infty}f_{N}(\beta)$.

From previous expressions of free energy density $f(\beta)$ a crucial consequence of the Bethe approximation on sparse graphs rises up: extensive quantities can be computed as a sum of local terms involving sites and edges of the graphs~\cite{Book_MezardMontanari2009}. Indeed, the same holds for internal energy density $e(\beta)=\partial_{\beta}(\beta f(\beta))$, while magnetizations can be easily computed from the marginal probability distributions $\eta_i(\theta_i)$. So in the end it is enough to solve self-consistency equations~(\ref{eq:self_cons_eta}) in order to be able to compute all the physical observables of the system.

So far, we have considered a given instance for the underlying RRG, and hence for the set of couplings $\{J_{ij}\}$. But when dealing with random topologies, in general, physical observables have to be computed by averaging over all the possible realizations of the graph and the disorder. This task can be accomplished by noting that cavity messages $\eta_{i\to j}$'s arriving from a branch of the tree are distributed according to a probability distribution $P[\eta_{i\to j}]$, and so BP equations~(\ref{eq:self_cons_eta}) can be reinterpreted as a single distributional equation:
\begin{multline}
P[\eta_{i\to j}]=\mathbb{E}_{G,J}\int\prod_{k=1}^{d_i-1}\mD\eta_{k\to i}\,P[\eta_{k\to i}]\\
\times\delta\Bigl[\eta_{i\to j}-\mF[\{\eta_{k\to i},J_{ik}\}]\Bigr],
\label{eq:self_cons_eta_distrib}
\end{multline}
where $\mathbb{E}_{G,J}$ stands for the expectation value over the realization of the graph and of the disorder, and $\mF$ is defined in Eq.~(\ref{eq:self_cons_eta}).
In particular, for the RRG ensemble, all the degrees $\{d_i\}$ are equal to $c$ and so the corresponding average can be ignored. Accordingly, the free energy density averaged over the RRG ensemble is given by
\begin{equation}
	f(\beta)=-\frac{1}{\beta}\,\mathbb{E}_{\eta,J}\Bigl[\ln{\mZ_i}\Bigr]+\frac{c}{2\beta}\,\mathbb{E}_{\eta,J}\Bigl[\ln{\mZ_{ij}}\Bigr]\,,
	\label{eq:f_rs_distrib}
\end{equation}
where the average over $\eta$ is made according to $P[\eta]$ satisfying Eq.~(\ref{eq:self_cons_eta_distrib}).

\subsection{Cavity equations at zero temperature}

When temperature $T$ goes to zero, the inverse temperature $\beta$ diverges and the integrals in Eq.~(\ref{eq:self_cons_eta}) can be solved by the saddle point method.
We rewrite the cavity marginals $\eta_{i\to j}$ as large deviation functions in $\beta$:
\begin{equation}
	\eta_{i\to j}(\theta_i)\equiv e^{\,\beta h_{i\to j}(\theta_i)}\,,
\end{equation}
with cavity fields $h_{i\to j}(\theta_i)$ being nonpositive functions.
The normalization on $\eta_{i\to j}$ requires to appropriately shift $h_{i\to j}$ such that its maximum has zero height:
\begin{equation}
	\max\nolimits_{\theta_i} h_{i\to j}(\theta_i) = 0\,.
	\label{eq:norm_h_T_0}
\end{equation}
In the $T\to 0$ limit, the BP equations become
\begin{equation}
\begin{split}
	h_{i\to j}(\theta_i) &= \mF_0[\{h_{k\to i}, J_{ik}\}]\\
	&= \sum_{k\in\partial i\setminus j}\max_{\theta_k}{\left[h_{k\to i}(\theta_k)+J_{ik}\cos{(\theta_i-\theta_k)}\right]}\,,
	\label{eq:self_cons_h_T_0}
\end{split}
\end{equation}
up to an additive constant due to the normalization condition in Eq.~(\ref{eq:norm_h_T_0}).

Taking the average over the disorder (couplings and RRGs of fixed degree $c$), we get again a self-consistency equation for functional probability distribution $P[h]$ of cavity fields $h$:
\begin{equation}
P[h]=\mathbb{E}_{J}\int \prod_{i=1}^{c-1} \mD h_i\,P[h_i]\,\delta\Bigl[h-\mF_0[\{h_i, J_i\}]\Bigr]\,.
	\label{eq:self_cons_h_T_0_distrib}
\end{equation}
The zero-temperature expression for the free energy density $f$ can be written in terms of the $P[h]$ satisfying Eq.~(\ref{eq:self_cons_h_T_0_distrib}),
\begin{equation}
	f=-\mathbb{E}_{h,J}\,\bigl[f_i\bigr]+\frac{c}{2}\,\mathbb{E}_{h,J}\,\bigl[f_{ij}\bigr]\,,
	\label{eq:f_rs_T_0_distrib}
\end{equation}
where
\begin{eqnarray*}
	f_i &=& \max_{\theta_i}{\biggl[\,\sum_{k\in\partial i}\max_{\theta_k}{\bigl[h_{k\to i}(\theta_k)+J_{ik}\cos{(\theta_i-\theta_k)}\bigr]}\biggr]}\,,\\
	f_{ij} &=& \max_{\theta_i ,\theta_j}{\Bigl[h_{i\to j}(\theta_i)+h_{j\to i}(\theta_j)+J_{ij}\cos{(\theta_i-\theta_j)}\Bigr]}\,.
\end{eqnarray*}

\subsection{Validity of the RS cavity method}
\label{subsec:validity_rs_method}

So far, we have implicitly assumed that the set of BP equations always admit a (unique) solution. However, in general, it is not always true, and so it is not obvious to get at a solution to eq.~(\ref{eq:self_cons_eta}). This fact is very intimately related to the presence of loops of finite size or strong correlations in the model.
Indeed, if the graph is a tree or locally tree-like, in the cavity graph where vertex $i$ has been removed the following factorization holds:
\begin{equation}
	\mu_{\partial i}\bigl(\{\theta_k\}_{k\in\partial i}\bigr)=\prod_{k\in\partial i}\eta_{k\to i}(\theta_k)\,,
	\label{eq:factor_mu}
\end{equation}
which in turn allows one to derive Eq.~(\ref{eq:self_cons_eta}).
But when the graph is not locally tree-like, i.\,e. it has short loops, or when correlations in the model are so strong that even in the cavity graphs without vertex $i$ the marginals $\eta_{k\to i}(\theta_k)$ are correlated, then factorization in Eq.~(\ref{eq:factor_mu}) does not hold anymore, and the error committed in assuming it can be non negligible, even in the case of very large graphs.

From the formal point of view, a rigorous proof of the conditions under which the replica symmetric cavity method is correct does not yet exist. A possible condition that has to be fulfilled regards the uniqueness of the Gibbs measure,
\begin{equation}
	\mu(\{\vs\})=\frac{1}{\mZ}e^{\,-\beta\mH[\{\vs\}]},
	\label{eq:gibbs_measure}
\end{equation}
meaning that the clustering property holds and that each spin $\vs_i$ in the bulk of the system is independent from any choice of boundary conditions. However, this is a very strict condition, and often it is observed that RS~cavity method still provides a correct result even when Gibbs measure~(\ref{eq:gibbs_measure}) ceases to be unique. So a weaker condition to be fulfilled regards the extremality of the Gibbs measure~(\ref{eq:gibbs_measure}), so that even when it is no longer unique but extremal, then in the thermodynamic limit the unique relevant solution is still the RS one~\cite{KrzakalaEtAl2007}. Roughly speaking, the extremality of the Gibbs measure means that the behavior of a spin $\vs_i$ in the bulk of the system depends only on a set of boundary conditions with null measure.

From the analytical and numerical point of view, instead, there are several and equivalent approaches for the study of the stability of the RS solution of a given model. For example, one can apply the 1RSB cavity method (Sec.~\ref{sec:bp_cavity_method_1rsb}) and check if it reduces to the RS solution, which is then exact in this case.

From a more physical point of view, one can compute the spin glass susceptibility~$\chi_{SG}$,
\begin{equation}
	\chi_{SG}=\frac{1}{N}\,\sum_{i,j}\braket{\vs_i\cdot\vs_j}^2_c,
\end{equation}
and see where it diverges, signaling a phase transition. 
The computation of $\chi_{SG}$ can be done in an iterative way via the so-called \emph{susceptibility propagation} algorithm, which in practice corresponds to check the growth rate of small perturbations around the fixed-point cavity messages: if these perturbations tend to grow, then the RS fixed point is no longer stable and the RS ansatz is only approximate. In Appendix~\ref{app:susc_propag} we deal with this topic in a deeper way, explicitly linearizing the BP equations --- both at finite and zero temperature --- and so obtaining the analytical expression of the growth rate of such perturbations.

The study of the growth rate of perturbations can be done either on \emph{given instances} either in \emph{population dynamics} taking the average over the ensemble of graphs and coupling realizations.
We exploit the second approach, redirecting the reader to Ref.~\cite{Thesis_Zdeborova2009} for the demonstration of the equivalence between them.

\section{RS solution of the XY model and the $q$-state clock model}
\label{sec:rs_solution_xy_clock}

Let us now apply the cavity method to the XY model and to the $q$-state clock model, in order to find their RS solution. We use first the bimodal distribution $J_{ij}=\pm 1$, with different weights $p$ and $1-p$, and we derive analytical phase diagrams in the~$p$ versus~$T$ plane.

We move then to the low-temperature region and numerically identify, when varying~$p$, three phases: the ferromagnetic (RS) one, the spin glass (RSB) one and also a \emph{mixed} phase between them, characterized by both a non vanishing magnetization and a breaking of the replica symmetry~\cite{CastellaniEtAl2005}.

Finally, we use a different disorder distribution, usually called \emph{gauge glass}, such that each interaction $J_{ij}$ acts through a rotation of a random angle $\omega_{ij}$, still belonging to one of the allowed directions of the $q$-state clock model. This further choice of disorder allows us to study the behavior of the clock model also for odd values of~$q$, since now there is no reflection symmetry to take care of. Furthermore, it should be more ``physical'' for the XY model, where in principle disorder can not only cause a complete inversion of a spin with respect to the direction it had in the ferromagnetic case, but also a small rotation.

\subsection{Analytical solution with the bimodal distribution of couplings}
\label{sec:analyticalBimodal}

Let us consider the following disorder distribution:
\begin{equation}
	\mathbb{P}_{J}(J_{ij})=p\,\delta(J_{ij}-1)+(1-p)\,\delta(J_{ij}+1)\;,
	\label{eq:disord_distr_pm_J}
\end{equation}
with $p\in[1/2,1]$, such that $p=1$ corresponds to a pure ferromagnet and $p=1/2$ to an unbiased spin glass.

Let us start our analytical computation from the XY~model. In order to find a solution to BP equations~(\ref{eq:self_cons_eta}) for the XY model, it is useful to expand cavity marginals $\eta_{k\to i}$ in Fourier series:
\[
\eta_{k\to i}(\theta_k)=
\frac{1}{2\pi}\biggl\{1+\sum_{l=1}^{\infty}\Bigl[a_{l}^{(k\to i)}\cos{(l\theta_k)}+b_{l}^{(k\to i)}\sin{(l\theta_k)}\Bigr]\biggr\},
\]
where Fourier coefficients are defined as usual:
\begin{equation}
	\left\{
	\begin{aligned}
	&a_l^{(i\to j)}=2\int d\theta_i\,\eta_{i\to j}(\theta_i)\,\cos{(l\theta_i)},\\
	&b_l^{(i\to j)}=2\int d\theta_i\,\eta_{i\to j}(\theta_i)\,\sin{(l\theta_i)}.
	\end{aligned}
	\right.
	\label{eq:def_fourier_coeff_XY}
\end{equation}
Note that, in general, Fourier coefficients are different for each cavity marginal.

Substituting this expansion in the right-hand side of BP equations~(\ref{eq:self_cons_eta}), we get
\begin{equation}
\begin{split}
	&\eta_{i\to j}(\theta_i)=\frac{1}{\mZ_{i\to j}}\,\prod_{k\in\partial i\setminus j}\int d\theta_k\,e^{\,\beta J_{ik}\cos{(\theta_i-\theta_k)}}\\
	&\,\,\,\,\frac{1}{2\pi}\biggl\{1+\sum_{l=1}^{\infty}\Bigl[a_{l}^{(k\to i)}\cos{(l\theta_k)}+b_{l}^{(k\to i)}\sin{(l\theta_k)}\Bigr]\biggr\},
	\label{eq:BPclock}
\end{split}
\end{equation}
\\\vspace{1.0cm}
where $\mZ_{i\to j}$ now reads
\begin{equation}
\begin{split}
	&{\mZ_{i\to j}} = \int d\theta_i \prod_{k\in\partial i\setminus j}\int d\theta_k\,e^{\,\beta J_{ik}\cos{(\theta_i-\theta_k)}}\\
	&\,\,\frac{1}{2\pi}\biggl\{1+\sum_{l=1}^{\infty}\Bigl[a_{l}^{(k\to i)}\cos{(l\theta_k)}+b_{l}^{(k\to i)}\sin{(l\theta_k)}\Bigr]\biggr\}\,.
	\label{eq:Zclock}
\end{split}
\end{equation}

Then, integrals in $d\theta_k$ can be performed by introducing modified Bessel functions of the first kind~\cite{Book_AbramowitzStegun1964}:
\begin{eqnarray*}
		&&\int d\theta_k\,e^{\,\beta J_{ik}\cos{(\theta_i-\theta_k)}}\cos{(l\theta_k)}=2\pi\,I_l(\beta J_{ik})\cos{(l\theta_i)}\,,\\
		&&\int d\theta_k\,e^{\,\beta J_{ik}\cos{(\theta_i-\theta_k)}}\sin{(l\theta_k)}=2\pi\,I_l(\beta J_{ik})\sin{(l\theta_i)}\,,\\
		&&\text{where} \quad I_n(x)\equiv\frac{1}{2\pi}\int_0^{2\pi}d\theta\,e^{x\cos{\theta}}\cos{(n\theta)}\,.
	\label{eq:modif_Bes_func_xy}
\end{eqnarray*}
BP equations thus become
\begin{equation*}
\begin{split}
	&\eta_{i\to j}(\theta_i)=\frac{1}{\mZ_{i\to j}}\,\prod_{k\in\partial i\setminus j}\biggl\{I_0(\beta J_{ik})\\
	&\quad +\sum_{l=1}^{\infty}I_l(\beta J_{ik})\Big[a_{l}^{(k\to i)}\cos{(l\theta_i)}+b_{l}^{(k\to i)}\sin{(l\theta_i)}\Bigr]\biggr\},
\end{split}
\end{equation*}
where also the normalization constant $\mZ_{i\to j}$ has to be rewritten in terms of Bessel functions by following the same steps. At this point, if we substitute this expression for $\eta_{i\to j}$ into~(\ref{eq:def_fourier_coeff_XY}), we get a set of self-consistency equations for the Fourier coefficients:
\begin{widetext}
\begin{equation}
	\left\{
	\begin{aligned}
	&a_{l}^{(i\to j)}=\frac{2}{\mZ_{i\to j}}\int d\theta\,\cos{(l\theta)}\prod_{k\in\partial i \setminus j} \biggl\{I_0(\beta J_{ik})+\sum_{p=1}^{\infty}I_p(\beta J_{ik})\Bigl[a_p^{(k\to i)}\cos{(p\,\theta)}+b_p^{(k\to i)}\sin{(p\,\theta)}\Bigr]\biggr\},\\
	&b_{l}^{(i\to j)}=\frac{2}{\mZ_{i\to j}}\int d\theta\,\sin{(l\theta)}\prod_{k\in\partial i \setminus j} \biggl\{I_0(\beta J_{ik})+\sum_{p=1}^{\infty}I_p(\beta J_{ik})\Bigl[a_p^{(k\to i)}\cos{(p\,\theta)}+b_p^{(k\to i)}\sin{(p\,\theta)}\Bigr]\biggr\},\\
	&\mZ_{i\to j} = \int d\theta\,\prod_{k\in\partial i \setminus j} \biggl\{I_0(\beta J_{ik})+\sum_{p=1}^{\infty}I_p(\beta J_{ik})\Bigl[a_p^{(k\to i)}\cos{(p\,\theta)}+b_p^{(k\to i)}\sin{(p\,\theta)}\Bigr]\biggr\}.
	\end{aligned}
	\right.
	\label{eq:self_cons_a_b}
\end{equation}
\end{widetext}

It is straightforward to verify that, in absence of any external field, the uniform distribution over the $[0,2\pi]$ interval is a solution of Eq.~(\ref{eq:self_cons_eta}) at any temperature:
\begin{equation}
	\eta_{i\to j}(\theta_i)=\frac{1}{2\pi}\quad,\qquad\forall i\to j,
	\label{eq:para_sol_xy_1}
\end{equation}
and obviously it corresponds to a vanishing solution for the self-consistency equations~(\ref{eq:self_cons_a_b}):
\begin{equation}
	a_{l}^{(i\to j)}=b_{l}^{(i\to j)}=0\quad,\qquad\forall i\to j\,,\,\forall l.
	\label{eq:para_sol_xy_2}
\end{equation}
This solution is nothing but the paramagnetic one, characterized by a set of null local magnetizations. Furthermore, it is very easy to compute the expression for the free energy density,
\begin{equation}
	f_{para}(\beta)=-\frac{1}{\beta}\ln{2\pi}-\frac{c}{2\beta}\ln{I_0(\beta)},
\end{equation}
as well as the expression for the internal energy density:
\begin{equation}
	e_{para}(\beta)=-\frac{c}{2}\,\frac{I_1(\beta)}{I_0(\beta)}.
\end{equation}

Assuming that, lowering the temperatures, a second-order phase transition takes place, we expect some of the Fourier coefficients to become nonzero in a continuous way.
To identify the critical line $T_c(p)$ we expand Eq.~(\ref{eq:self_cons_a_b}) to linear order in the Fourier coefficients. At linear order, the normalization constant is just
\[
\mZ_{i\to j} = 2\pi\,\prod_{k\in\partial i\setminus j}I_0(\beta J_{ik}).
\]
Expanding also numerators in Eq.~(\ref{eq:self_cons_a_b}), and restricting to $a$'s coefficients (since expressions for $b$'s coefficients are similar), we get
\begin{equation}
	a_l^{(i\to j)}=\sum_{k\in\partial i\setminus j}\frac{I_l(\beta J_{ik})}{I_0(\beta J_{ik})}a_l^{(k\to i)}
	\label{eq:self_cons_a_linear}
\end{equation}
and, taking the average over the disorder distribution and the graph realization, one gets a self-consistency equation for the mean values of Fourier coefficients,
\begin{equation}
	\overline{a_l} = \mathbb{E}_{G,J}\,\left[a_l^{(i\to j)}\right],
\end{equation}
that depends on the parity of the coefficient, namely
\begin{eqnarray}
	\overline{a_l} = (c-1)(2p-1)\frac{I_l(\beta)}{I_0(\beta)} \overline{a_l}
	&&\qquad \text{for $l$ odd},\\
	\overline{a_l} = (c-1)\frac{I_l(\beta)}{I_0(\beta)} \overline{a_l}
	&&\qquad \text{for $l$ even}.
\label{mom1}
\end{eqnarray}
For $\beta>0$, the ratios $I_l(\beta)/I_0(\beta)$ are increasing functions of $\beta$.
Moreover, the inequality $I_{l+1}(\beta) < I_l(\beta)$ implies that the first 
Fourier coefficients to become nonzero lowering the temperature are
$a_1$ (for~$p$ close to~$1$) and $a_2$ (for~$p$ close to~$1/2$).

However, we have to consider that Fourier coefficients are random variables fluctuating from edge to edge. In strongly disordered models ($p$ close to $1/2$), the mean value of $a_1$ may stay zero, while fluctuations may become relevant and eventually diverge. To check for this, we compute the self-consistency equation for the second moments:
\begin{equation}
\overline{a_l^2} =\mathbb{E}_{G,J}\,\left[\Bigl(a_l^{(i\to j)}\Bigr)^2\right] =(c-1)\frac{I_l^2(\beta)}{I_0^2(\beta)}\,\overline{a_l^2}.
\label{mom2}
\end{equation}
The comparison of Eqs.~(\ref{mom1}) and~(\ref{mom2}), together with the inequality
\begin{equation}
	\frac{I_2(x)}{I_0(x)} \le \frac{I_1^2(x)}{I_0^2(x)}\,,
\end{equation}
lead to the conclusion that the instability of paramagnetic phase, lowering the temperature, is always driven by the instability in the first order Fourier coefficients~$\{a_1^{(i\to j)}\}$.
The low-temperature phase will be ferromagnetic in case the stability produces a nonzero mean value~$\overline{a_1}$, which in turn leads to a nonzero magnetization. While a spin glass order prevails if the transition is such that $\overline{a_1^2}$ becomes nonzero, while~$\overline{a_1}$ stays null.
Which phase transition takes actually place depends on the highest critical temperature between $T_F=1/\beta_F$ and $T_{SG}=1/\beta_{SG}$, with
\begin{eqnarray}
	(c-1)(2p-1)\frac{I_1(\beta_F)}{I_0(\beta_F)} &=& 1,\\
	(c-1)\frac{I_1^2(\beta_{SG})}{I_0^2(\beta_{SG})} &=& 1.
\end{eqnarray}
These results completely agree with those obtained for the XY model on Erd\H{o}s-R\'enyi graphs in Refs.~\cite{SkantzosEtAl2005,CoolenEtAl2005} through a slightly different approach, namely a functional moment expansion around paramagnetic solution.

Notice that $T_F$ does depend on density of ferromagnetic couplings $p$, while $T_{SG}$ does not.
The \emph{multicritical} point~$(p_{mc},T_{mc})$ is located where $T_F(p)$ and $T_{SG}$ meet, i.\,e.
\begin{equation}
	p_{mc}=\frac{1+(c-1)^{-1/2}}{2} \qquad , \quad T_{mc} = T_{SG}\,.
	\label{eq:abscissa_p_mc}
\end{equation}
Just below the critical temperature
\[
T_c=\max\big(T_F(p),T_{SG}\big)
\]
the nonlinear terms in Eq.~(\ref{eq:self_cons_a_b}) that couple different Fourier coefficients lead to the following scaling
\begin{equation}
	a_l^{(i\to j)} \propto \tau^{\,l/2}
	\label{eq:scalingCoeff}
\end{equation}
with $\tau \equiv |T-T_c|/T_c$.
Equation~(\ref{eq:scalingCoeff}) can be obtained noticing the following two aspects of Eq.~(\ref{eq:self_cons_a_b}): \textit{i)} for $l>1$ the first nonlinear term is
\[
a_l^{(i\to j)} \propto \prod_{k\in\partial i\setminus j} a_{p_k}^{(k \to i)},
\]
with $p_k$ algebraically summing to $l$, implying $a_l \propto (a_1)^l$;
\textit{ii)} the first nonlinear term in the equation for $a_1$ is cubic and thus, close to the critical point, we have
\[
a_1 = (1+\tau) a_1 + A\, a_1^3
\]
implying $a_1 \propto \sqrt{\tau}$.

The above Fourier expansion can be used only for identifying the instability of the paramagnetic phase in absence of an external field. Unfortunately, in presence of a field or in the low-temperature phases all the Fourier coefficients become $O(1)$ and the above expansion becomes useless, since keeping few coefficients is a too drastic approximation.

So, in order to complete the phase diagram and locate the critical line between the spin glass and the ferromagnetic phases that runs from $(p_{mc},T_{mc})$ to $(p_{SG},T=0)$, we will move to a numerical approach based on the RS cavity method (see Sec.~\ref{subsec:numerical_cavity_method}).

Before moving to the low-temperature phase, let us conclude the analysis of the critical lines of the paramagnetic phase in the $q$-state clock model, following exactly the same Fourier expansion above.
Since our aim is to understand how fast the clock model converges to the XY model increasing $q$, we would like to study a model where the $q$ dependence is smooth.
Using the bimodal distribution for couplings, $J_{ij}=\pm1$, we are forced to work only with even values of $q$. Indeed, for odd values of $q$, couplings with $J_{ij}=1$ can be fully satisfied, while coupling with $J_{ij}=-1$ can not (there are no two states differing by a $\pi$ angle for $q$ odd).
Although for large $q$ the differences between even and odd $q$ models vanish, for small $q$ values they lead to strong oscillations in most physical observables. For this reason we focus only on even values for $q$ as long as we use bimodal couplings.

First of all, let us rewrite BP equations for the~$q$-state clock model. They are still given by Eq.~(\ref{eq:self_cons_eta}), with a slight modification due to the discrete nature of the model:
\begin{equation}
\begin{aligned}
	\eta_{i\to j}(\theta_{i,a})&=\frac{1}{\mZ_{i\to j}}\,\prod_{k\in\partial i\setminus j}\,\sum_{b_k=0}^{q-1}\eta_{k\to i}(\theta_{k,b_k})\\
	&\qquad\qquad\qquad\times e^{\,\beta J_{ik}\cos{\left(\theta_{i,a}-\theta_{k,b_k}\right)}},
	\label{eq:self_cons_eta_qclock}
\end{aligned}
\end{equation}
where indices $a$ and $b_k$'s label the $q$ possible values for angles $\theta$'s, and $\mZ_{i\to j}$ is given by
\begin{equation*}
	\mZ_{i\to j}=\frac{2\pi}{q}\sum_{a=0}^{q-1}\prod_{k\in\partial i\setminus j}\sum_{b_k=0}^{q-1}\eta_{k\to i}(\theta_{k,b_k})\,
e^{\,\beta J_{ik}\cos{\left(\theta_{i,a}-\theta_{k,b_k}\right)}}.
\end{equation*}

Since we are now dealing with a discrete model, in order to find a solution to BP equations~(\ref{eq:self_cons_eta_qclock}) it is useful to expand cavity marginals $\eta_{i\to j}$ in discrete Fourier series,
\begin{equation}
	\eta_{i\to j}(\theta_{i,a})=\frac{1}{q}\sum_{b=0}^{q-1}c_b^{(i\to j)}e^{\,2\pi i\,ab/q},
	\label{eq:discr_four_series}
\end{equation}
where the complex coefficients $c_b^{(i\to j)}$ are given by
\begin{equation}
	c_b^{(i\to j)}=\sum_{a=0}^{q-1}\eta_{i\to j}(\theta_{i,a})\,e^{-2\pi i\,ab/q}.
\end{equation}
The zero-order coefficient $c_0^{(i\to j)}$ is nothing but the sum of the values taken by cavity marginal $\eta_{i\to j}$ over the~$q$~values of the angle $\theta_i$:
\begin{equation}
	c_0^{(i\to j)}=\sum_{a=0}^{q-1}\eta_{i\to j}(\theta_{i,a}).
\end{equation}
If we choose to put the norm of probability distributions for the $q$-state clock model equal to $q/2\pi$, so that in the limit $q\to\infty$ we can exactly recover the marginal probability distributions for the XY model, then we have $c_0^{(i\to j)}=q/2\pi$ and we can write
\begin{equation}
	\eta_{i\to j}(\theta_{i,a})=\frac{1}{2\pi}\biggl[1+\frac{2\pi}{q}\,\sum_{b=1}^{q-1}c_b^{(i\to j)}e^{\,2\pi i\,ab/q}\biggr].
\end{equation}
A different choice is to put the norm equal to $1$, so obtaining $c_0=1$. This choice gives the correct value for physical observables for a discrete model (e.\,g. an entropy which is always positive defined), but it will be less useful when studying the convergence of the $q$-state clock model toward the XY model. So from now on, we will always use the $q/2\pi$ normalization.

It is worth noticing that
\begin{equation}
	c_m^{(i\to j)}=\left(c_{q-m}^{(i\to j)}\right)^*\,,
\end{equation}
given that cavity messages $\eta_{i\to j}$ are real quantities. In particular $c_{q/2}$ is real, since we are using $q$ even.

Expanding in discrete Fourier series both sides of BP equations~(\ref{eq:self_cons_eta_qclock}),
we get the following self-consistency equations for the Fourier coefficients:
\begin{multline}
c_{m}^{(i\to j)}=\frac{1}{\mZ_{i\to j}}\,\sum_{a=0}^{q-1} e^{-2\pi i\,am/q}\\
\prod_{k\in\partial i\setminus j}\,\sum_{b_k=0}^{q-1} c_{b_k}^{(k\to i)}\,I_{b_k}^{(q)}(\beta J_{ik})\,e^{\,2\pi i\,ab_k/q},
	\label{eq:self_cons_c_qclock}
\end{multline}
where also $\mZ_{i\to j}$ has to be expressed in terms of discrete Fourier coefficients.
In order to keep a compact notation we have introduced the discrete analogous of modified Bessel functions of the first kind,
\begin{equation}
	I_n^{(q)}(x) \equiv \frac{1}{q}\sum_{a=0}^{q-1}\,e^{\,x\cos(2\pi a/q)}\cos\bigg(\frac{2\pi n a}{q}\bigg)\,,\quad n\in\mathbb{Z},
	\label{eq:modif_Bes_func_qclock}
\end{equation}
which converge to usual Bessel functions in the large $q$ limit:
\[
\lim_{q\to\infty} I_n^{(q)}(x) = I_n(x)\,.
\]

In analogy with the XY model, also BP equations~(\ref{eq:self_cons_eta_qclock}) admit the paramagnetic solution, given by the uniform distribution over the $q$ values (note the $q/2\pi$ normalization)
\begin{equation}
	\eta_{i\to j}(\theta_{i,a})=\frac{1}{2\pi}\quad,\quad\forall i\to j\,,\,\forall a\in\{1,\dots,q-1\}
	\label{eq:para_sol_qclock_1}
\end{equation}
that corresponds to a vanishing solution for the self-consistency equations: (\ref{eq:self_cons_c_qclock})
\begin{equation}
	c_{m}^{(i\to j)}=0\quad,\qquad\forall i\to j\,,\,\forall m\in\{1,\dots,q-1\}.
	\label{eq:para_sol_qclock_2}
\end{equation}
The corresponding expressions for free energy density $f^{(q)}(\beta)$ and internal energy density $e^{(q)}(\beta)$ are
\begin{equation}
	f_{para}^{(q)}(\beta)=-\frac{1}{\beta}\ln{2\pi}-\frac{c}{2\beta}\ln{I_0^{(q)}(\beta)},
\end{equation}
\begin{equation}
	e_{para}^{(q)}(\beta)=-\frac{c}{2}\,\frac{I_1^{(q)}(\beta)}{I_0^{(q)}(\beta)}.
\end{equation}
As expected, thank to the choice of the $q/2\pi$ normalization, these expressions converge to those for the XY model in the $q\to\infty$ limit.

The next step is to study the stability of paramagnetic solution.
An analysis analogous to the one made for the XY model tells us that most unstable coefficient are first order ones, satisfying at linear order the following equations
\begin{equation}
	c_{1}^{(i\to j)}=\sum_{k\in\partial i\setminus j}\frac{I_1^{(q)}(\beta J_{ik})}{I_0^{(q)}(\beta J_{ik})}\,c_{1}^{(k\to i)}
	\label{eq:self_cons_c_1_qclock}
\end{equation}
By averaging over the disorder distribution~$\mathbb{P}_J(J_{ij})$, we get self-consistency equations for the first two momenta,
\begin{equation}
\left\{
\begin{aligned}
	&\overline{c_1}=(c-1)(2p-1)\frac{I_1^{(q)}(\beta)}{I_0^{(q)}(\beta)}\,\overline{c_1},\\
	&\overline{c_1^2}=(c-1)\,\left[\frac{I_1^{(q)}(\beta)}{I_0^{(q)}(\beta)}\right]^2\,\overline{c_1^2},
\end{aligned}
\right.
\end{equation}
that identify critical temperatures for the phase transitions towards a ferromagnet, $T_F^{(q)}(p)=1/\beta_F^{(q)}(p)$, and towards a spin glass, $T_{SG}^{(q)}=1/\beta_{SG}^{(q)}$:
\begin{equation}
	\left\{
	\begin{aligned}
	&(c-1)(2p-1)\frac{I_1^{(q)}\big(\beta_F^{(q)}(p)\big)}{I_0^{(q)}\big(\beta_F^{(q)}(p)\big)}=1,\\
	&(c-1)\,\left[\frac{I_1^{(q)}\big(\beta_{SG}^{(q)}\big)}{I_0^{(q)}\big(\beta_{SG}^{(q)}\big)}\right]^2=1.
	\end{aligned}
	\right.
\end{equation}
The paramagnetic phase is stable for temperatures larger than the critical one:
\[
T_c^{(q)}=\max\big(T_F^{(q)}(p),T_{SG}^{(q)}\big)\,.
\]
In analogy with the XY model, for $p$ close to $1$, the clock model has a transition towards a ferromagnetic phase, while for $p$ close to $1/2$ the transition is towards a spin glass phase.
Surprisingly, the abscissa $p_{mc}$ of the multicritical point in the $q$-state clock model has exactly the same expression in Eq.~(\ref{eq:abscissa_p_mc}) found for the XY model.

\begin{figure}
	\centering
	\includegraphics[width=\columnwidth]{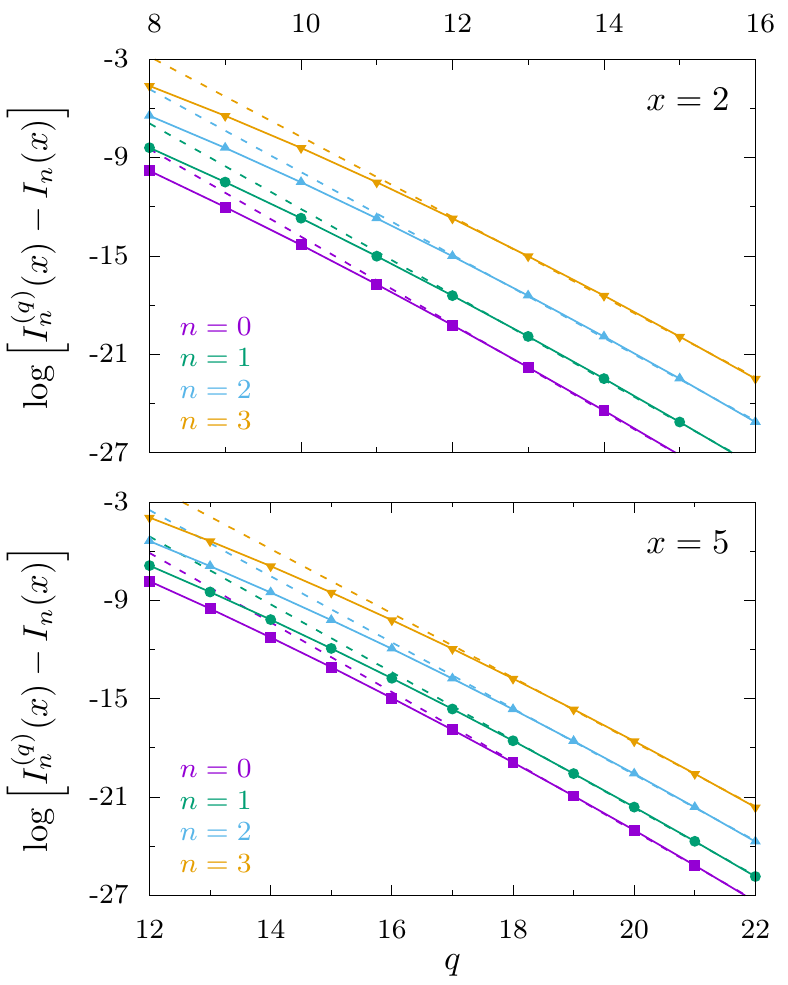}
	\caption{Convergence of the discretized modified Bessel functions $I_n^{(q)}(x)$ toward their limiting values when $q\to\infty$, computed at $x=5$ (upper dataset) and $x=2$ (lower dataset). For each value of $n$, we plot the logarithm of $I_n^{(q)}(x)-I_n(x)$ together with a linear fit, to highlight the exponential convergence in $q$.}
	\label{fig:conv_Bessel_log}
\end{figure}

The only dependence of these critical lines on the number $q$ of states is through the discrete Bessel function $I_n^{(q)}$. So in order to understand how fast the clock model phase diagram converges to the one of the XY model, we need to study the rate of convergence of the functions $I_n^{(q)}(x)$ to the Bessel functions $I_n(x)$ in the large $q$ limit.
We show in Fig.~\ref{fig:conv_Bessel_log} a numerical evidence that this convergence is exponentially fast in $q$, i.\,e. like $\exp(-q/q^*)$, with a characteristic scale $q^*$ increasing with the argument $x$:
\begin{align*}
	q^*(x=2) &\simeq 2.0\,,\\
	q^*(x=5) &\simeq 2.5\,.
\end{align*}
Numerical evidence shows that $q^*$ is finite for any finite value $x$, but it seems to diverge in the $x\to\infty$ limit. In that limit, the convergence may follow a stretched exponential.

\begin{figure*}[!t]
	\centering
	\includegraphics[width=\linewidth]{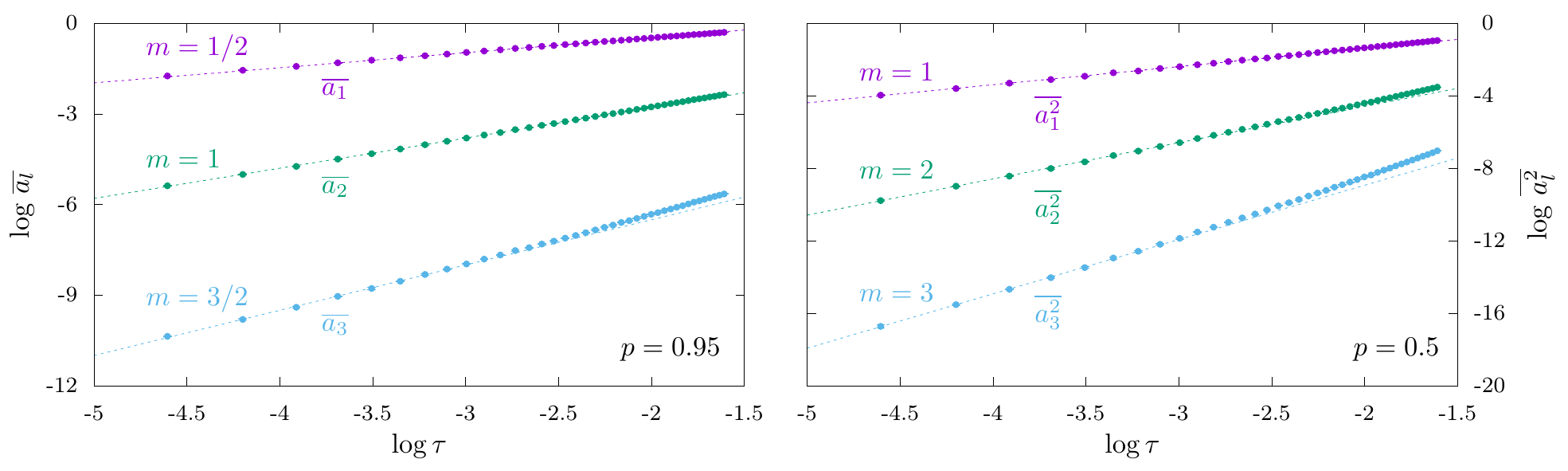}
	\caption{The first momenta of the lowest-order Fourier coefficients just below the critical temperature in the $q=64$ clock model (a very good approximation to the XY model), measured via population dynamics ($\mN=10^5$). In the left panel we use $p=0.95$ (para-ferro transition) and plot first moment. In the right panel $p=0.5$ (para-spin glass transition) and we plot second moment (the first one being null). The slope of each line, $m$, is reported. Data follows the analytical expectation $a_l \propto \tau^{l/2}$, with statistical errors smaller than the symbol size.}
	\label{fig:Fourier_coefficients}
\end{figure*}

Unfortunately, we have not been able to find a fully analytical proof of this statement.
The following argument should, however, convince the reader that a power law decay in $q$ is not expected to take place every time one approximates the integral of a periodic function with a finite sum of $q$ terms.
Let us suppose $f(x)$ is an infinitely differentiable function, $2\pi$-periodic, i.\,e. $f(x+2\pi)=f(x)$, and we are interested in approximating the integral
\[
I(f) = \frac{1}{2\pi} \int_0^{2\pi} dx\,f(x)
\]
with the finite sum
\[
I^{(q)}(f) = \frac{1}{q} \sum_{a=0}^{q-1} f(2\pi a/q).
\]
Rewriting $I^{(q)}$ as the integral of a step-wise function, the error $\Delta^{(q)}=I^{(q)}-I$ can be written as the sum of $q$ local terms, each one computed in a small interval $\Gamma_a\equiv [2\pi a/q-\pi/q,2\pi a/q+\pi/q]$ of size $2\pi/q$ around $2\pi a/q$:
\begin{equation*}
\Delta^{(q)}(f) = \frac{1}{2\pi} \sum_{a=0}^{q-1}\,\,\int_{\Gamma_a}dx\,\bigl[f(x)-f(2\pi a/q)\bigr].
\end{equation*}
For large $q$, we can Taylor expand the integrand around the central point of each interval $\Gamma_a$,
\[
f(x) - f(2\pi a/q) = \sum_{k=1}^\infty f^{(k)}(2\pi a/q) \frac{(x-2\pi a/q)^k}{k!}\,,
\]
where $f^{(k)}$ is the $k$th derivative of $f$. Thus the error is given by the following series:
\[
\Delta^{(q)}(f) = \sum_{\substack{k \text{ even}\\k>0}} \frac{\pi^k}{q^{k+1} (k+1)!} \sum_{a=0}^{q-1} f^{(k)}(2\pi a/q).
\]
For $q$ large, the internal sum can be approximated by the $q\to\infty$ limit, plus the error term,
\begin{multline*}
\frac1q \sum_{a=0}^{q-1} f^{(k)}(2\pi a/q) 
= \frac{1}{2\pi}\int_0^{2\pi}dx\,f^{(k)}(x) + \Delta^{(q)}(f^{(k)})=\\
\,\,\, = \frac{f^{(k-1)}(2\pi) - f^{(k-1)}(0)}{2\pi}+ \Delta^{(q)}(f^{(k)}) = \Delta^{(q)}(f^{(k)})\,,
\end{multline*}
where the last inequality follows from the $2\pi$ periodicity.
So, the equation for the error term is given by
\begin{equation}
\Delta^{(q)}(f) = \sum_{\substack{k \text{ even}\\k>0}} \frac{\pi^k}{q^k (k+1)!} \,\Delta^{(q)}(f^{(k)})\,.
\label{eq:error}
\end{equation}
For a function $f$ smooth enough --- like the one in the definition of the modified Bessel functions of first kind, $f(\theta)=\exp[x \cos(\theta)]\cos(n\theta)$ --- we expect the error on the derivatives, $\Delta^{(q)}(f^{(k)})$, to decay with $q$ in the same way as the error on the function itself, $\Delta^{(q)}(f)$.
This expectation is further confirmed by the data in Fig.~\ref{fig:conv_Bessel_log}, where we see that the error on the function $I_0$ decays as the error on its derivative $I_1$.

Noticing that the power-law ansatz $\Delta^{(q)}(f^{(k)}) \propto q^{-\alpha}$ is incompatible with Eq.~(\ref{eq:error}) for any value of the power~$\alpha$, we conclude that the error $\Delta^{(q)}(f)$ decays faster that any power law.

Apart from excluding a power law decay, the above argument is not able to provide the final answer: e.\,g. whether the decay is a simple exponential decay or a stretched one. The evidence presented in Sec.~\ref{sec:convergence} will suggest the decay is exponential for any positive temperature and changes to a stretched exponential at $T=0$.

\subsection{Numerical solution with the bimodal distribution of couplings}
\label{subsec:numerical_cavity_method}

As already explained above, the analytical expansion in Fourier series can be used only in the high temperature phase. The low-temperature region can be fully explored and understood only by using numerical tools. In particular, we will implement the cavity method at the RS stage, both at finite and zero temperature, by exploiting the \emph{population dynamics algorithm}. This method, firstly introduced in Ref.~\cite{AbouChacraEtAl1973} and then revisited and refined in Refs.~\cite{MezardParisi2001, MezardParisi2003}, allows one to compute physical observables averaged over the disorder distribution and the graph realizations.

To this purpose, we consider a population of $\mN$ cavity marginals $\eta(\theta)$, randomly initialized, that evolve according to the iterative BP equations: at each step of the algorithm, each marginal in the population is updated according to the following equation:
\[
\eta_\ell(\theta) \leftarrow \mF\big(\{\eta_{i_k}, J_k\}_{k=1,\ldots,c-1}\big)\,,
\]
where $\mF$ is defined in Eq.~(\ref{eq:self_cons_eta}), $J_k$'s are random variables generated according to the coupling distribution $\mathbb{P}_J$, and $i_k$'s are random indices uniformly drawn in $[1,\mN]$, so as to choose $c-1$ random marginals in the population.

\begin{figure*}[t]
	\centering
	\includegraphics[width=\linewidth]{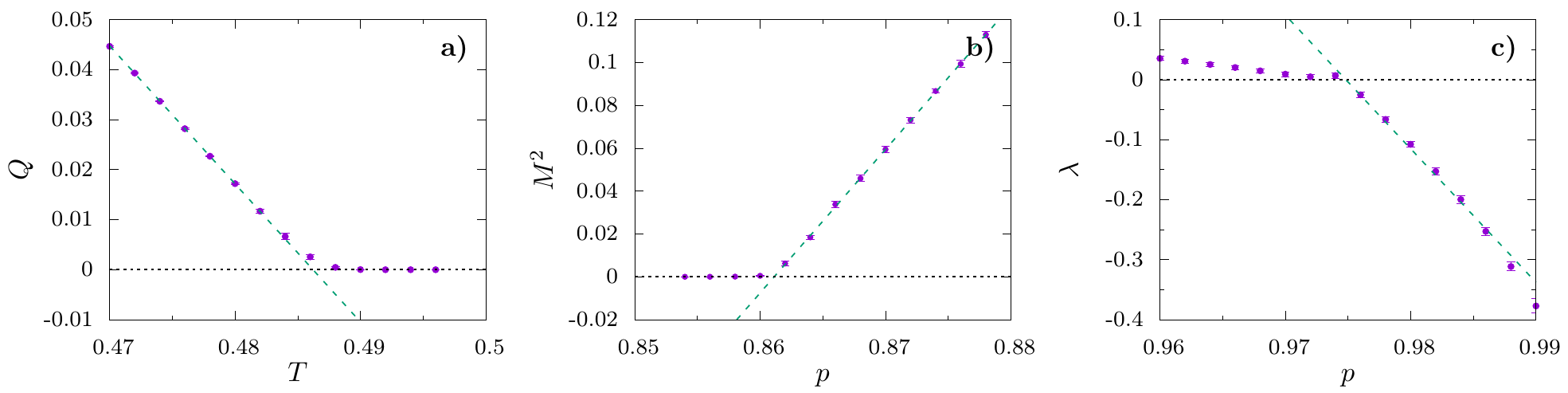}
	\caption{Computation of the different critical lines in the $p$ vs $T$ phase diagram for the $8$-state clock model. \textbf{a)} The overlap $Q(T)$ with $p=0.5$ signals the paramagnetic to spin glass phase transition at $T_c(p=0.5)=0.4862(1)$. \textbf{b)} The square of total magnetization $M^2(p)$ with $T=0.2$ marks the transition between mixed and spin glass phases at $p_c(T=0.2)=0.8611(1)$. \textbf{c)} The growing rate of perturbations $\lambda(p)$ with $T=0.2$ becomes positive at the phase transition between ferromagnetic and mixed phases at $p_*(T=0.2)=0.9748(2)$.}
	\label{fig:critical_lines_J_Q008}
\end{figure*}

Physical observables, which are functionals of the marginals in the population, usually change during the first part of the evolution, and then converge to an asymptotic value, corresponding to the thermodynamical expectation of that observable (within the replica symmetric ansatz).
Being the algorithm of a stochastic nature and the population of finite size, we expect fluctuations of $O(1/\sqrt{\mN})$. In what follows, if not stated otherwise, we use a population of $\mN=10^6$ cavity marginals and a fixed degree $c=3$ for the underlying RRG.
We have also checked that estimates of physical observables are compatible with what can be measured on a given samples of large size; however, the population dynamics algorithm is more efficient in computing physical observables averaged over the RRG ensemble and coupling distribution.

At variance with the BP equations~(\ref{eq:self_cons_eta}) that may not have a solution, the population dynamics algorithm always converges to a fixed-point probability distribution of marginals $\mP^*[\eta]$, independently from the initial conditions. Furthermore, this is true even when the RS assumption is no longer correct: when this happens, the distribution $\mP^*[\eta]$ we get is no longer the exact one, and so we have to use (at least) the 1RSB ansatz. This will be done in Sec.~\ref{sec:bp_cavity_method_1rsb}.

Before searching for the transition lines between the different low-temperature phases in the $q$-state clock model and the XY model, we would like to verify the scaling of Fourier coefficients just below the critical temperature. In Fig.~\ref{fig:Fourier_coefficients}, we report the results of this check both for the para-ferro phase transition ($p=0.95$, left panel) and for the para-spin glass phase transition ($p=0.5$, right panel). In the former case we have $\overline{a_l} \propto \tau^{l/2}$, while in the latter 
$\overline{a_l} = 0$ and $\overline{a_l^2} \propto \tau^{l}$, as expected from the computation in Sec.~\ref{sec:analyticalBimodal}.

In the low-temperature phases, each spin variable has a nonzero average:
\begin{equation}
	\vec{m}_i = (m_{i,x}, m_{i,y}) = \braket{\,(\cos{(\theta_i)},\sin{(\theta_i)})\,}\,,
\end{equation}
where the angular brackets represent the average over a full marginal $\eta_i(\theta_i)$ defined in Eq.~(\ref{eq:fullMarginals}).
From the local magnetizations $\vec{m}_i$'s we can build two order parameters: the norm of global magnetization vector,
\begin{equation}
	M \equiv \left| \frac1N \sum_i \vec{m}_i \right|\,,
\end{equation}
and the overlap,
\begin{equation}
	Q \equiv \frac{1}{N} \sum_i |\vec{m}_i|^2\,,
\end{equation}
which satisfy the inequality $M^2\le Q$.
From the analysis of Fourier coefficients shown before, we expect both~$Q$ and~$M^2$ to grow linearly below the critical temperature~$T_c$ (see Fig.~\ref{fig:critical_lines_J_Q008}, left panel).

In the paramagnetic phase, all local magnetizations are null ($M=Q=0$), while in a pure ferromagnetic phase, all spins are perfectly aligned and so $M^2=Q>0$.
In the more general case ($p<1$ and $T<T_c$), local magnetizations exist, but do not align perfectly and so we have $Q>0$ and $M^2<Q$: the unbiased spin glass phase ($M=0$, $Q>0$) belongs to this class, but also two other phases --- the disordered ferromagnet and the magnetized spin glass, the so-called mixed phase --- have $0<M^2<Q$ and can not be distinguished by just looking at these two order parameters. In order to distinguish this two phases we will need to check for the stability of the RS solution with respect to a breaking of the replica symmetry. This scenario is very similar to the one taking place in the Ising model~\cite{CastellaniEtAl2005}.

As long as we keep track only of the order parameters $Q$ and $M$, in the low-temperature phase ($T<T_c$), we have that $Q>0$ anywhere, while $M$ is non zero only for $p$ large enough. In central panel of Fig.~\ref{fig:critical_lines_J_Q008}, we show the typical behavior of $M^2$ as a function of $p$ at $T=0.2<T_c$ for the $q=8$ clock model.
The critical $p_c$ estimated this way is an approximation to the true critical line separating the unbiased spin glass phase and the mixed phase; the right computation should be done within a full replica symmetry breaking ansatz, which is unfortunately unavailable for the diluted models we are studying here.
We expect, however, the RS ansatz to provide a very good approximation.

The critical line separating the RSB mixed phase from the RS disordered ferromagnet can be computed by studying the stability of the RS fixed point via the \emph{susceptibility propagation} (SuscP) algorithm, that we run in population dynamics.

As explained in Sec.~\ref{subsec:validity_rs_method} and in Appendix~\ref{app:susc_propag}, the SuscP algorithm amounts at studying the stability of the linearized BP equations around the RS solution.
At any finite temperature, the perturbations around the fixed-point cavity marginals $\eta^*_{i\to j}$ evolve via the following linearized equations:
\begin{equation}
	\delta\eta_{i\to j} = \sum_{k\in\partial i \setminus j}\,\Biggl|\frac{\delta\,\mF[\{\eta\}]}{\delta\,\eta_{k\to i}}\Biggr|_{\eta^*_{k\to i}}\delta\eta_{k\to i}\,,
\end{equation}
where $\mF$ is defined in Eq.~(\ref{eq:self_cons_eta}).
We check for the growth of these perturbations by measuring the following norm:
\begin{equation}
	|\delta\eta|\equiv\sum_{(i\to j)}\,\sum_{a=0}^{q-1}\,|\delta\eta_{i\to j}(\theta_{i,a})|\,.
\end{equation}
We define the growing rate as
\begin{equation}
\lambda = \lim_{t\to\infty} \frac{\ln|\delta\eta|}{t}\,.
\end{equation}
In the right panel of Fig.~\ref{fig:critical_lines_J_Q008}, we report the values of $\lambda$ as a function of $p$, for $T=0.2<T_c$ in the $q=8$ clock model. The $p_*$ value where $\lambda=0$ corresponds to the phase transition between a mixed RSB phase and an RS disordered ferromagnet: it is the point where the spin glass susceptibility diverges \cite{parisi2014diluted}.
In the RSB phase the perturbations growing rate $\lambda$ is strictly positive, because the RS solution is unstable there.

\begin{figure}[t]
	\centering
	\includegraphics[width=\columnwidth]{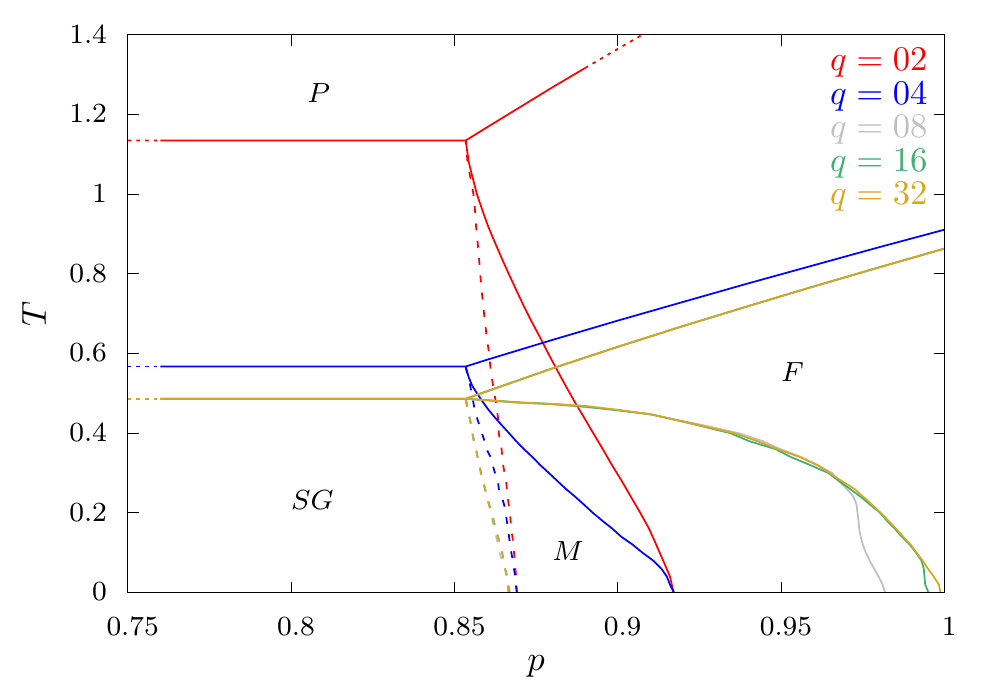}
	\caption{Phase diagram $p$ versus $T$ of the $q$-state clock model with bimodal coupling and different values of $q$. Full critical lines are exact, while dashed ones are approximated. The convergence to the XY model in the $q\to\infty$ limit is very fast and critical lines with $q \ge 8$ are practically superimposed, but for the region in the lower right corner.}
	\label{fig:phase_diag_J}
\end{figure}

Having explained the way we compute the different critical lines, we can now draw in Fig.~\ref{fig:phase_diag_J} the full phase diagram for the $q$-state clock model, with several values of~$q$.
We notice that the convergence to the XY model in the $q\to\infty$ limit is very fast: in practice, critical lines with $q\ge 8$ are superimposed and coincide with those in the XY model.
The only region where a $q$ dependence is still visible is that with $p$ close to~$1$ and $T$ close to~$0$.
In this region we have a strong dependence on the discretization, such that for small~$q$ values the RS disordered ferromagnet ($p$ is close to 1, but strictly smaller) is stable down to $T=0$, while in the XY model ($q\to\infty$ limit) there is always a phase transition to an RSB phase lowering the temperature with $p<1$.

This is an important new finding (to the best of our knowledge it was not known before). It suggests the XY model may show RSB effects much more easily than the Ising model, when the disorder is weak.
The reason for this behavior is maybe due to the fact that, in presence of a weak disorder, which is not strong enough to force discrete variables in different directions, the continuous variables in the XY model can adapt more easily to several different orientations (states).

An analogous behavior when increasing $q$ is also found in the study of the $q$-state clock model in the $d=3$ cubic lattice by means of Migdal-Kadanoff approximate renormalization group~\cite{IlkerBerker2013}. Again, paramagnetic-ferromagnetic critical line converges very fast in $q$, while a stronger dependence in $q$ is found for the ferromagnetic-spin glass critical line (moving toward larger fractions of ferromagnetic couplings, as in our case) and the paramagnetic-spin glass critical line (moving toward the zero-temperature axis, unlike our case).

\subsection{Numerical solution of the gauge glass model}

In order to discuss the gauge glass model, it is convenient to rewrite the Hamiltonian in a different form:
\begin{equation}
	\mathcal{H}[\{\vs\}]=-\sum_{\braket{ij}}\cos{(\theta_i-\theta_j-\omega_{ij})}\,,
	\label{eq:H_xy_gg}
\end{equation}
where $\omega_{ij}$ are the preferred relative orientation between neighboring spins.

The model with bimodal couplings studied above corresponds to $\omega_{ij} \in \{0,\pi\}$.
In the gauge glass model, instead, the random rotations $\omega_{ij}$ take values uniformly in $[0,2\pi)$.
The latter choice seems more in line with the continuous nature of the variables.

The straightforward extension to the $q$-state clock model suggests to take $\omega_{ij} \in \{0,2\pi/q,\ldots,2\pi(q-1)/q\}$.
Willing to interpolate with a single parameter between the pure ferromagnetic model ($\omega_{ij}=0$) and the unbiased spin glass model ($\omega_{ij}$ uniformly distributed), we choose the following coupling distribution:
\begin{equation}
	\mathbb{P}_{\omega}^{(q)}(\omega_{ij})=(1-\Delta)\,\delta(\omega_{ij})+\frac{\Delta}{q}\sum_{a=0}^{q-1}
	\delta\left(\omega_{ij}-\frac{2\pi\,a}{q}\right)
	\label{eq:disord_distr_gg}
\end{equation}
with $\Delta\in[0,1]$. In this way, when $q\to\infty$ we exactly recover the uniform continuous distribution for the XY gauge glass model. Furthermore, in this model we can use any $q$ value, since we do not expect any difference between even and odd values: any pair of spins can in principle assume a configuration satisfying a coupling~$\omega_{ij}$ for any $q$~value.

\begin{figure}[t]
	\centering
	\includegraphics[width=\columnwidth]{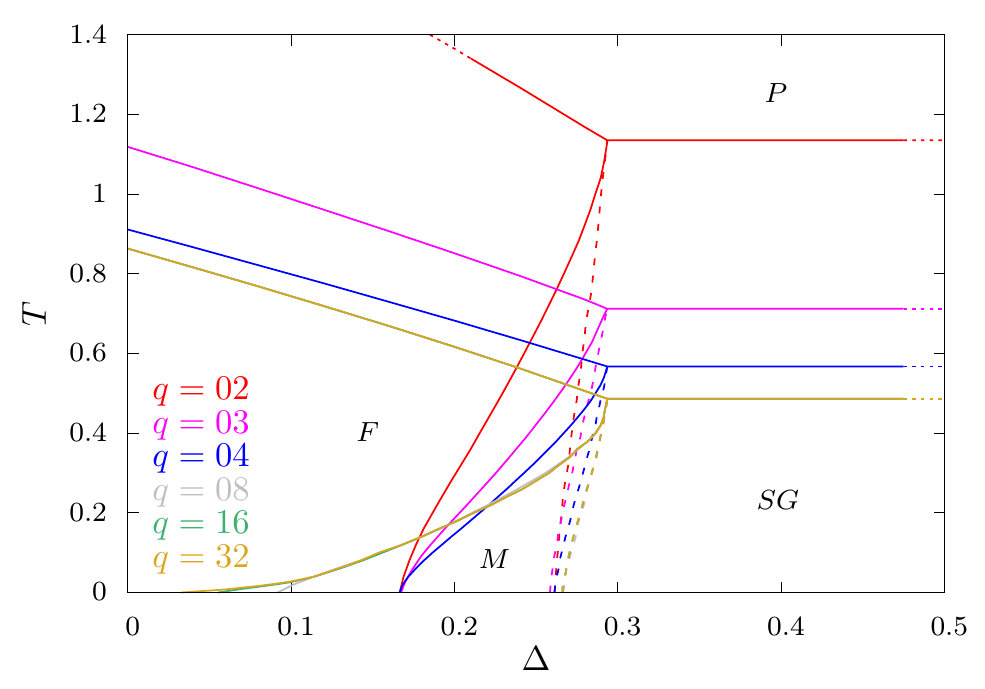}
	\caption{Phase diagram $\Delta$ versus $T$ of the $q$-state clock model with the gauge glass couplings and different values of $q$. Full critical lines are exact, while dashed ones are approximated. The convergence to the XY model in the $q\to\infty$ limit is very fast and critical lines with $q \ge 8$ are practically superimposed, but for the region in the lower left corner.}
	\label{fig:phase_diag_W}
\end{figure}

\begin{figure*}[t]
	\centering
	\includegraphics[width=\linewidth]{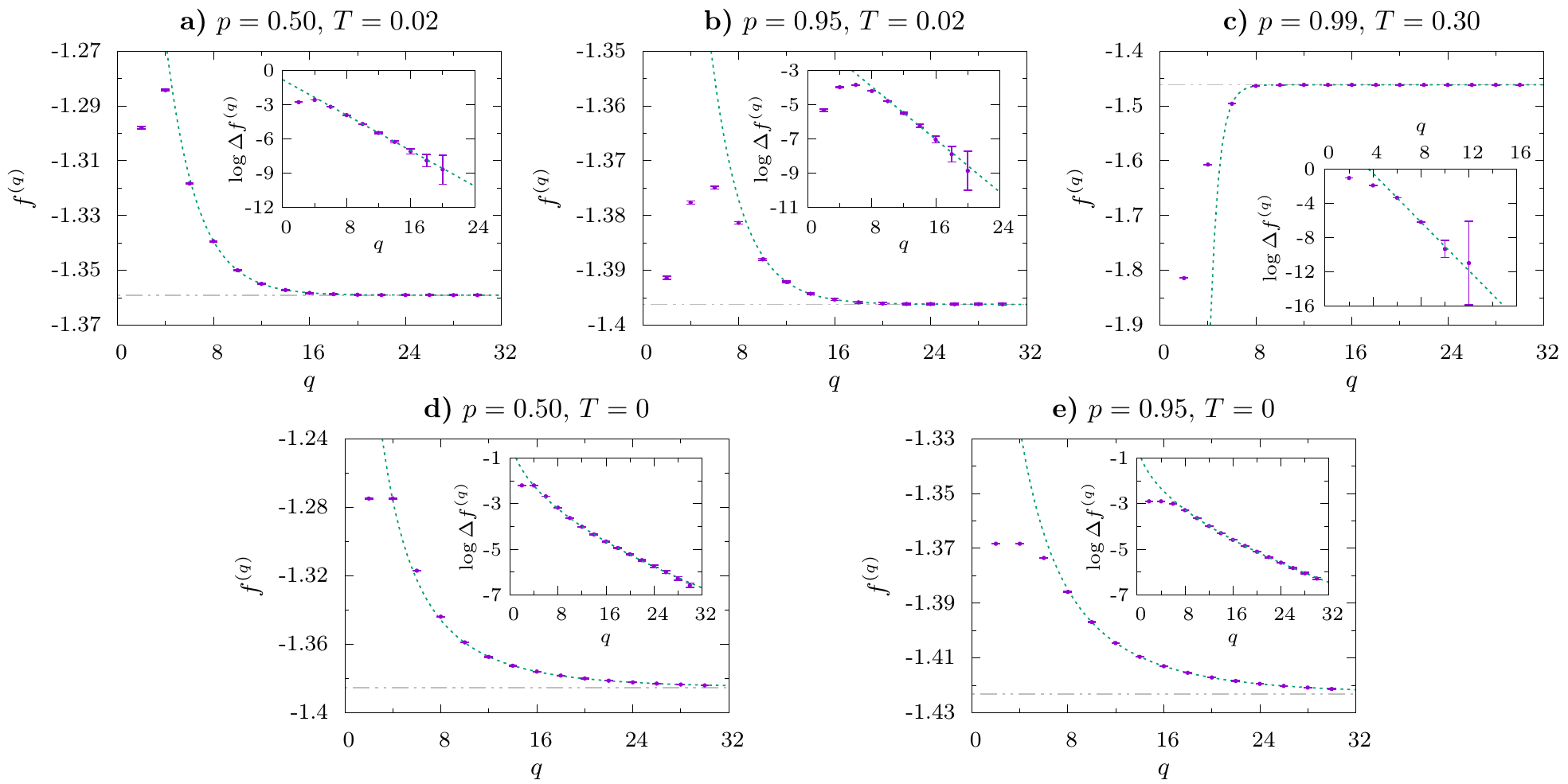}
	\caption{Convergence of the free energy density $f^{(q)}(\beta)$ of the $q$-state clock model toward that of the XY model for the $J_{ij}=\pm 1$ disorder distribution. Panels \textbf{a)}, \textbf{b)} and \textbf{c)} show the convergence at finite temperature in the unbiased spin glass, mixed and ferromagnetic phases, respectively. The fits correspond to an exponential decay. Panels \textbf{d)} and \textbf{e)} show the convergence at zero temperature in the unbiased spin glass and mixed phase. In this case, the fits are stretched exponential functions.}
	\label{fig:multiplot_freeEnergy_J_N1_00e+07}
\end{figure*}

By using the same techniques exposed above, we derive the phase diagram of this model as a function of~$T$ and~$\Delta$. The results are shown in Fig.~\ref{fig:phase_diag_W}, where we draw critical lines for several different values of $q$. As in the bimodal case, the convergence to the XY model in the $q\to\infty$ limit is very fast and already for $q=8$ most of the phase diagram does not depend on $q$ anymore and provides the result for the XY model.
Also in this case the only region where the $q$~dependence is stronger is the one where the temperature is close to zero and the disorder is very weak.
In the $q\to\infty$ limit, the XY model seems again to show RSB effects for any infinitesimal amount of disorder in the $T=0$ limit.

We notice \textit{en passant} that for $q=2$ the two versions of the clock model are identical up to the transformation $p \leftrightarrow 1-\Delta/2$. This is clearly visible in Figs.~\ref{fig:phase_diag_J} and~\ref{fig:phase_diag_W}, where the red lines corresponding to the $q=2$ clock model are identical up to a horizontal reflection.

\section{Convergence of physical observables}
\label{sec:convergence}

We face now the main task of this work, studying the convergence of physical observables of the $q$-state clock model toward those of the XY model. We measure physical observables via the population dynamics algorithm, but we need to use a population size $\mN=10^7$ in order to achieve the required accuracy.

The results presented in the previous sections about the fast convergence of the phase diagrams and the exponential convergence of discretized Bessel functions, strongly suggest an exponential convergence of physical observables as long as $T>0$.
Indeed, as shown by the three upper panels in Fig.~\ref{fig:multiplot_freeEnergy_J_N1_00e+07}, the free-energy of the clock model converges to the one of the XY model exponentially fast in $q$ as long as $T>0$.
In these plots, we restrict the analysis to the low-temperature phases (spin glass, mixed, and ferromagnetic), because in the paramagnetic phase the convergence is so fast that it is hardly measurable.

\begin{figure*}[t]
	\centering
	\includegraphics[width=\linewidth]{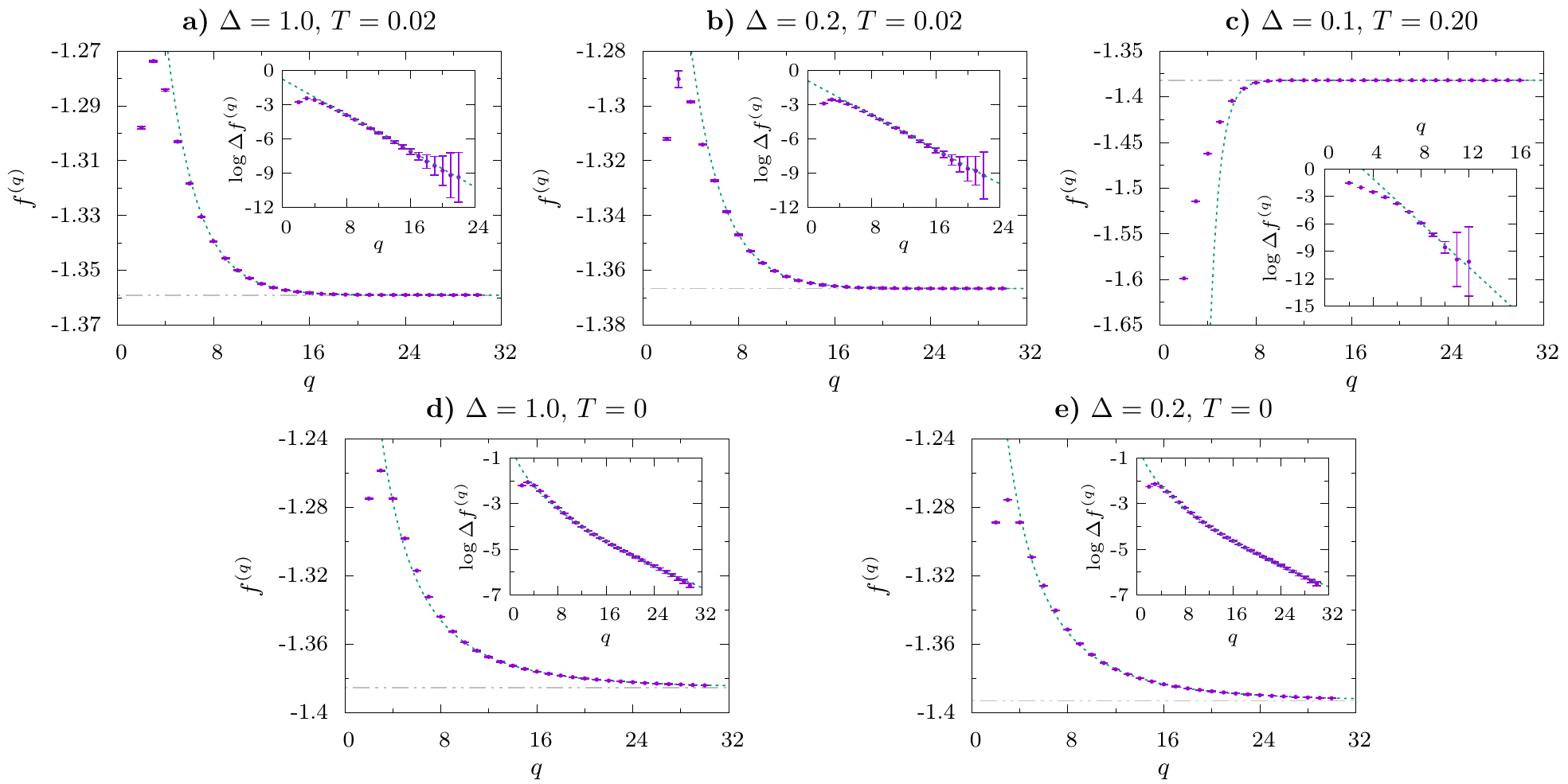}
	\caption{Convergence of the free energy density $f^{(q)}(\beta)$ of the $q$-state clock model toward that of the XY model for the gauge glass disorder distribution, Eq.~(\ref{eq:disord_distr_gg}). Panels \textbf{a)}, \textbf{b)} and \textbf{c)} show the convergence at finite temperature in the unbiased spin glass, mixed and ferromagnetic phases, respectively. The fits correspond to an exponential decay. Panels \textbf{d)} and \textbf{e)} show the convergence at zero temperature in the unbiased spin glass and mixed phases. In this case, the fits are stretched exponential functions.}
	\label{fig:multiplot_freeEnergy_W_N1_00e+07}
\end{figure*}

We fit data at $T>0$ via the exponential function
\[
	\ln{\Delta f^{(q)}(T)} \equiv \ln{\left[f^{(q)}(T)-f^{(\infty)}(T)\right]}= A-q/q^* 
\]
estimating the following values for $q^*$ (all fits have an acceptable $\chi^2$ per degree of freedom, as shown by the values reported below on the right):
\begin{equation*}
\begin{gathered}
	q^*(p=0.50,\,T=0.02)=2.57(1)\,, \qquad \chi^2/dof=0.20/5\,,\\
	q^*(p=0.95,\,T=0.02)=2.60(5)\,, \qquad \chi^2/dof=0.19/3\,,\\
	q^*(p=0.99,\,T=0.30)=0.70(1)\,, \qquad \chi^2/dof=0.12/2\,.
\end{gathered}
\end{equation*}
As soon as $q \gg q^*$ the clock model provides an extremely good approximation to the XY model physical observables, with a systematic error which is by far much smaller than the typical statistical uncertainty achieved in numerical simulations. In this sense, a Monte Carlo study of the clock model with $q$ large enough can be a much more efficient way of measuring physical observables in the XY model.

At $T=0$, we observe a slower convergence in $q$, which is well fitted by a stretched exponential
\begin{equation}
	\ln{\Delta f^{(q)}(T=0)} = A-(q/q^*)^b\,.
\end{equation}
For the two cases reported in Fig.~\ref{fig:multiplot_freeEnergy_J_N1_00e+07}, we find that the $b$ exponent is very close to $1/2$ (and thus we fix it to that value in the fits), while values for $q^*$ are the following:
\begin{equation*}
\begin{gathered}
	q^*(p=0.50)=0.67(1)\,, \qquad \chi^2/dof=6.4/9\,,\\
	q^*(p=0.95)=0.79(1)\,, \qquad \chi^2/dof=2.4/9\,.
\end{gathered}
\end{equation*}
Although these values are slightly smaller than those in the $T>0$ case, the $b\simeq 1/2$ exponent makes the convergence at $T=0$ slower.

So, it seems that even in the slowest case ($T=0$) the convergence of clock model observables to those of the XY model is fast enough to safely allow to use the clock model with moderately large values of $q$.

In Fig.~\ref{fig:multiplot_freeEnergy_W_N1_00e+07}, we report the analogous results for the clock model with the gauge glass coupling distribution in Eq.~(\ref{eq:disord_distr_gg}). Also in this case, convergence in $q$ is exponentially fast for $T>0$ with the following parameters:
\begin{equation*}
\begin{gathered}
	q^*(\Delta=1.0,\,T=0.02)=2.55(1)\,, \qquad \chi^2/dof=0.48/12\,,\\
	q^*(\Delta=0.2,\,T=0.02)=2.66(1)\,, \qquad \chi^2/dof=1.35/13\,,\\
	q^*(\Delta=0.1,\,T=0.20)=0.83(1)\,, \qquad \chi^2/dof=0.43/4\,.
\end{gathered}
\end{equation*}
It is interesting to note that these values are similar to the ones found by using the bimodal coupling distribution.

Again, at $T=0$, the convergence becomes a stretched exponential, with a $b$ exponent close to $1/2$ for any $\Delta$ value:
\begin{equation*}
\begin{gathered}
	q^*(\Delta=1.0)=0.67(1)\,, \qquad \chi^2/dof=12.5/19\,,\\
	q^*(\Delta=0.2)=0.68(1)\,, \qquad \chi^2/dof=12.8/19\,.
\end{gathered}
\end{equation*}

\section{1RSB Cavity Method}
\label{sec:bp_cavity_method_1rsb}

We have shown that in both the XY model and the $q$-state clock model at low temperatures the replica symmetric ansatz breaks down if the disorder is strong enough (low $p$ or large $\Delta$ values).
We observe this replica symmetry breaking (RSB) via the RS fixed point becoming unstable.

In the RSB phase, we know that the RS result is just an approximation, although we expect it to be a rather good approximation for some observables (e.\,g. self-averaging observables, like the energy).
Nonetheless, in order to keep track of the RSB effects and the many states present in a RSB phase, we can resort to a more complicated ansatz with one step of replica symmetry breaking~(1RSB).

In models with pairwise interactions, like those we are studying here, we expect a full RSB ansatz to be required in the strongly disordered and low-temperature phase. Nonetheless, even the 1RSB results can be illuminating on the true physical behavior.
We thus solve the $q$-state clock model by means of the 1RSB cavity method derived by M\'ezard and Parisi~\cite{MezardParisi2001}.

The presence of many states breaks the validity of factorization in Eq.~(\ref{eq:factor_mu}). Indeed, by adding nodes and links to the graph by following the RS cavity method prescriptions within each state, one realizes that each state gets a different free-energy shift and thus this leads to a reweighing of the different states \cite{MezardParisi2001}.

We redirect the reader to book~\cite{Book_MezardMontanari2009} and lecture notes~\cite{Zamponi2008} for an exhaustive description of the 1RSB cavity method for solving sparse disordered models.
Here we just sketch the key concepts involved in the 1RSB solution.

At a given temperature $T=1/\beta$, the number of states with free energy density $f$ in a system of size $N$ can be written as
\begin{equation}
	\mN(f) = e^{\,N\Sigma_\beta(f)}\,,
\end{equation}
where $\Sigma_\beta(f)$ is called \emph{complexity} in the literature on spin glasses and \emph{configurational entropy} in that on structural glasses.
Introducing a \emph{replicated partition function}
\begin{equation}
\begin{split}
	\mZ_\beta(x) \equiv \sum_\alpha e^{-\beta N x f_\alpha} &\simeq \int df \,e^{\,N[\Sigma_\beta(f) - \beta x f]}\\
	&\simeq e^{\,N\big[\Sigma_\beta\big(f^*(\beta,x)\big)-\beta x f^*(\beta,x)\big]}\,,
\end{split}
\end{equation}
where $f_\alpha$ is the free energy of state $\alpha$ and $f^*(\beta,x)$ is the maximizer of $\Sigma(f) - \beta x f$, that depends on both $\beta$ and $x$, one can compute $\Sigma(f)$ as the Legendre transform of the \emph{replicated free energy} $\phi_\beta(x)$:
\begin{eqnarray*}
	\phi_\beta(x) &\equiv& -\,\frac{1}{\beta x N} \ln\mZ_\beta(x)\,,\\
	f^*(\beta,x) &=& \phi_\beta(x) + x\,\partial_x\phi_\beta(x)\,,\\
	\Sigma(\beta,x) &=& \beta x^2\,\partial_x\phi_\beta(x)\,.
\end{eqnarray*}
The complexity $\Sigma(f)$ can be obtained by plotting parametrically $\Sigma(\beta,x)$ versus $f^*(\beta,x)$ varying $x$ at fixed $\beta$.

Thermodynamical quantities are obtained by setting $x=1$ if the corresponding complexity is positive, i.\,e. $\Sigma(\beta,x=1)\ge 0$. Otherwise, if $\Sigma(\beta,x=1)<0$, these states do not exists, and the partition function is dominated by the states with $x=x^*<1$ such that $\Sigma(\beta,x=x^*)=0$, where $x^*$ is called Parisi parameter.

\subsection{1RSB equations and their solution by means of population dynamics algorithm}

The computation of the replicated partition function $\mZ_\beta(x)$ must take into account the presence of many states, each one weighed by $\exp(-\beta x f_\alpha)$.

In each state $\alpha$ BP equations~(\ref{eq:self_cons_eta}) are still valid; let us refer to them briefly as $\eta_{i\to j}=\mF[\{\eta_{k\to i}\}]$. Since now we have to reweigh the cavity messages $\eta$ according to the free energy shift they produce, it is necessary to introduce a probability distribution $\mP[\cdot]$ over the RS probability distribution $P[\eta]$ of cavity messages. These two levels of populations come from the two different averages that we have to perform in the 1RSB approach: \textit{i)} a first average in a given state, that gives the RS population $P[\eta]$, and \textit{ii)} a second average over the states, that gives the 1RSB population $\mP[P]$.

If $\mZ_{i\to j}$ is the normalization constant that comes from the computation of cavity message $\eta_{i\to j}$ via the RS BP equation $\eta_{i\to j}=\mF[\{\eta_{k\to i}\}]$, i.\,e. the free energy shift due to the addition of site $i$ and directed edges $(k\to i)$'s with $k\in\partial i \setminus j$, then the reweigh acts as follows
\begin{equation}
\begin{split}
	P_{i\to j}[\eta_{i\to j}] &\equiv \mG[\{P_{k\to i}\}]\\
	&=\mathbb{E}_{G,J}\int\prod_{k\in\partial i \setminus j}\mD\eta_{k\to i}\,P_{k\to i}[\eta_{k\to i}]\quad\quad\\
	&\qquad\times\delta\Bigl[\eta_{i\to j}-\mF[\{\eta_{k\to i}\}]\Bigr]\Bigl(\mZ_{i\to j}[\{\eta_{k\to i}\}]\Bigr)^x
	\label{eq:self_cons_eta_distrib_reweigh}
\end{split}
\end{equation}
In this way the RS solutions are reweighed by $\exp{(-\beta N m f_{\alpha})}$, as required in the computation of $\mZ(x)$. The average over all the states yields the distributional equation for probability distribution $\mP[P]$:
\begin{equation}
	\mP[P_{i\to j}]=\int \prod_{k\in\partial i \setminus j}\mD P_{k\to i}\,\mP[P_{k\to i}]
  \delta\Bigl[P_{i\to j}-\mG[\{P_{k\to i}\}]\Bigr]
	\label{eq:self_cons_eta_distrib_1rsb}
\end{equation}
Eqs.~(\ref{eq:self_cons_eta_distrib_reweigh}) and~(\ref{eq:self_cons_eta_distrib_1rsb}) are solved by a population dynamics algorithm, that considers the two levels of average. We store $\mN$ populations each one made of~$\mM$ cavity messages~$\eta$. Cavity messages evolve via Eq.~(\ref{eq:self_cons_eta_distrib_reweigh}), where populations entering the r.\,h.\,s. are randomly chosen according to Eq.~(\ref{eq:self_cons_eta_distrib_1rsb}). For each population, the reweighing of messages is performed as explained in Ref.~\cite{Book_MezardMontanari2009}, by first computing $r\mM$ (with $r>1$) new cavity messages, and then selecting $\mM$ among these with a probability proportional to $\mZ^x$.
Typical values for the $r$ parameter are contained in the range $[2,5]$.

Physical observables in the thermodynamical limit can be written as averages over the two levels of probability distributions \cite{MezardParisi2001,Book_MezardMontanari2009}, namely over populations $P$'s and over cavity marginals $\eta$'s:
\begin{eqnarray*}
	\phi_\beta(x) &=& \frac{c}{2\beta x}\,\mathbb{E}_P\Bigl[
			\ln{\mathbb{E}_{\eta}\left[\mZ_{ij}^x\right]}
		\Bigr] - \frac{1}{\beta x}\,\mathbb{E}_P\Bigl[
			\ln{\mathbb{E}_{\eta}\left[\mZ_i^x\right]}
		\Bigr]\,,\\
f(\beta,x) &=& -\frac{1}{\beta}\,\mathbb{E}_P\left[
			\frac{\mathbb{E}_{\eta}\left[\mZ_i^x \ln{\mZ_i}\right]}
			{\mathbb{E}_{\eta}\left[\mZ_i^x\right]} \right] +\\
			&& + \frac{c}{2\beta}\,\mathbb{E}_P\left[
			\frac{\mathbb{E}_{\eta}\left[\mZ_{ij}^x \ln{\mZ_{ij}}\right]}
			{\mathbb{E}_{\eta}\left[\mZ_{ij}^x\right]}\right]\,,\\
\Sigma(\beta,x) &=& \beta x \Big(f(\beta,x) - \phi_\beta(x)\Big)\,.
\end{eqnarray*}

\begin{figure*}[t]
	\centering
	\includegraphics[width=\linewidth]{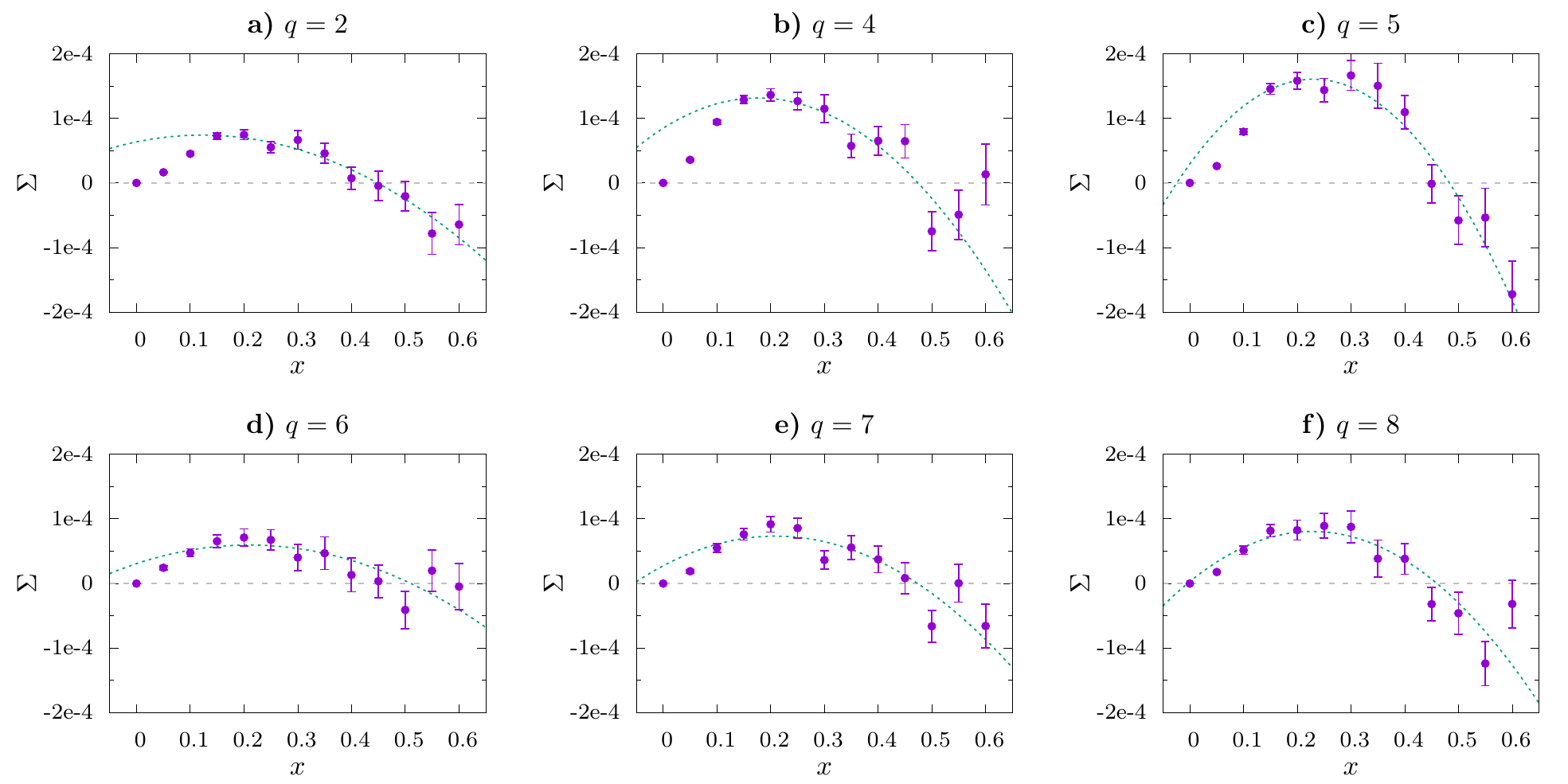}
	\caption{Plot of complexity $\Sigma(x)$ for the $q$-state clock model with bimodal couplings ($p=1/2$) at reduced temperature $\tau=1/2$, i.\,e. $T=T_c/2$. Plot ranges are the same in all the panels. We also draw the fitting quadratic curve used for each $q$~value to estimate the Parisi parameter $x^*$, such that $\Sigma(x^*)=0$.}
	\label{fig:sigma_1rsb}
\end{figure*}

In the 1RSB ansatz, thanks to the two levels of average, we can define two different overlaps:
an inner overlap $Q_1$, describing the similarity of local magnetizations inside a given state:
\begin{equation}
	Q_1 = \mathbb{E}_P\left[
					\frac{\mathbb{E}_{\eta}\left[\mZ_i^x\bigl(m_{i,x}^2+m_{i,y}^2\bigr)\right]}
					{\mathbb{E}_{\eta}\left[\mZ_i^x\right]}
			   \right]\,,
\end{equation}
and an outer overlap $Q_0$, describing the similarity of magnetizations between different states:
\begin{equation}
Q_0 = \mathbb{E}_P\left[
						\frac{\mathbb{E}_{\eta}\bigl[\mZ_i^x m_{i,x}\bigr]}
						{\mathbb{E}_{\eta}\bigl[\mZ_i^x\bigr]}
				   \right]^2
	+\mathbb{E}_P\left[
						\frac{\mathbb{E}_{\eta}\bigl[\mZ_i^x m_{i,y}\bigr]}
						{\mathbb{E}_{\eta}\bigl[\mZ_i^x\bigr]}
					\right]^2\,.
\end{equation}
As expected, it holds $Q_1 \geqslant Q_0$. These two overlaps are nothing but the analogous of the ones introduced in the 1RSB solution to the SK model by Parisi~\cite{Parisi1980a,Parisi1980b}. We are going to use these overlaps to approximate the Parisi function, $Q(x)$, in the spin glass phase, given that the full RSB solution is not known for disordered models on sparse graphs.

\subsection{1RSB solution of the $q$-state clock model}

The 1RSB ansatz is known to provide the correct solution to many disordered models defined on random sparse graphs (e.\,g. $p$-spin models~\cite{MezardEtAl2003}, random $K$-SAT problems~\cite{MontanariEtAl2004} and random coloring problems~\cite{ZdeborovaKrzakala2007}), at least in a range of parameters.
These models have either interactions involving $p>2$ variables ($p$-spin and $K$-SAT models) or variables taking $q>2$ values (coloring problems).

For the $q$-state clock model on a sparse random graph (hereafter we restrict to symmetric bimodal couplings, $p=1/2$), it is less clear how close to the right solution the 1RSB ansatz is.
We expect a continuous phase transition for most $q$ values, and so the 1RSB solution should be seen as an approximation to the correct full RSB solution.
Our expectation comes from the following observations.
For $q=2$, the clock model coincides with an Ising model, and for $q=4$, it is equivalent to a double Ising model.
For $q=3$, the clock model is equivalent to a \mbox{$3$-state} Potts model or $q=3$ coloring problem that has no thermodynamical phase transition on a random 3-regular graph~\cite{KrzakalaEtAl2004}, but only a dynamical phase transition is expected to happen (well described by a 1RSB ansatz). This case can be studied by fixing $x^*=1$ as in Ref.~\cite{MarruzzoLeuzzi2016}, where the $q$-state clock model with multi-body interactions is studied in the 1RSB frame.
For $q\ge5$, there are no known results on sparse graphs and our results in Sec.~\ref{sec:analyticalBimodal} suggest a continuous transition for any $q$ value.
This may look at variance with results for the $q$-state Potts model, but in the clock model the $q$ states have a specific ordering that eventually converge to the orientations of a continuous variable in the $q\to\infty$ limit.
Given that in that limit the transition is again continuous, it is reasonable to expect that the transition in the disordered $q$-state clock model on a sparse random graph is continuous for any $q$ values.

The literature provides one more piece of evidence in this direction.
Nobre and Sherrington~\cite{NobreSherrington1986} have studied the $q$-state clock model on the complete graph, finding a continuous phase transition for any $q \neq 3$ value.
Moreover, they have expanded the replicated free energy close to the critical point, i.\,e. for $\tau=(T_c-T)/T_c\ll 1$, where the Parisi function $Q(x)$ can be well approximated by a linear function $Q(x)=a\,x$ for $x<b\,\tau$ and a constant $Q(x)=a\,b\,\tau$ for $x\ge b\,\tau$. The parameters $a$ and $b$ determine the universality class. Nobre and Sherrington~\cite{NobreSherrington1986} found that $q=2$ and $q=4$ belong to the Ising universality class, while $q=3$ is in the $3$-state Potts one and for $q\ge 5$ the universality class is the same as the one of the XY model.

\begin{figure*}[t]
	\centering
	\includegraphics[width=\linewidth]{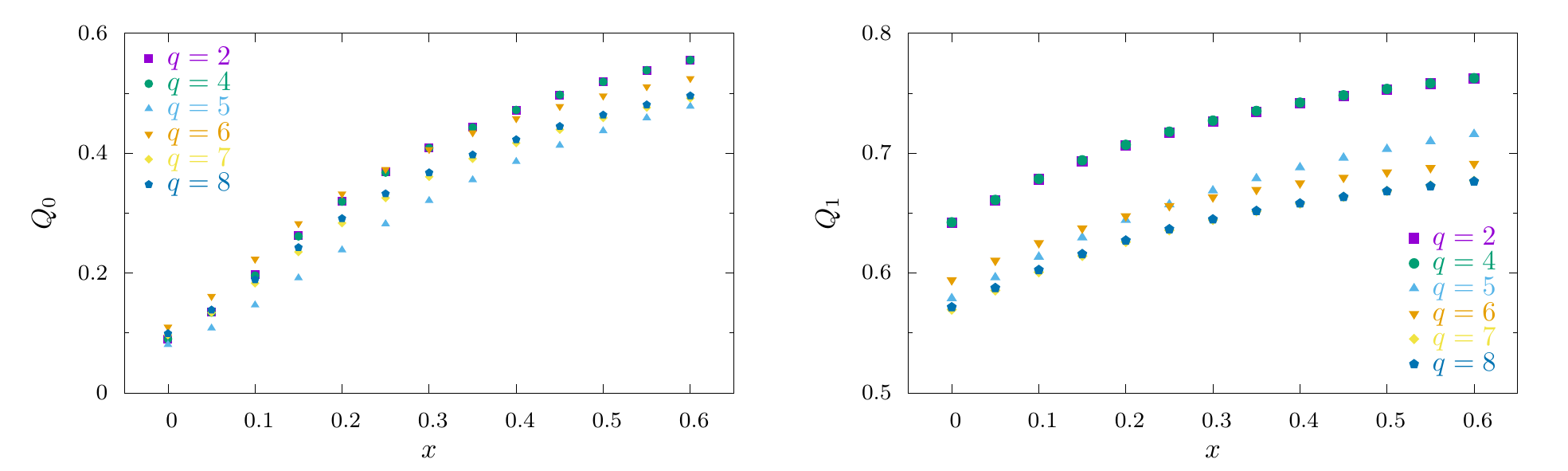}
	\caption{Plots of overlaps $Q_1$ and $Q_0$ in the 1RSB solution for the $q$-state clock model with bimodal couplings ($p=1/2$) at reduced temperature $\tau=1/2$, i.\,e. $T=T_c/2$. Please notice the different ranges on the $y$ axis for the two plots.}
	\label{fig:overlap_1rsb}
\end{figure*}

Unfortunately, for diluted models with a continuous transition, the study of the 1RSB solution very close to the critical point is infeasible, because the replicated free energy $\phi_\beta(x)$ differs from the constant RS free energy $\phi_\beta(0)$ by a quantity going to zero linearly in $\tau$, and --- given the complexity is very small (see Fig.~\ref{fig:sigma_1rsb}) --- computing numerically $\phi_\beta(x)$ at very small $\tau$ is too noisy.
For this reason we have computed $\phi_\beta(x)$ at $\tau=1/2$, namely in the middle of the spin glass phase.

We focus on values of $q$ ranging up to $8$, excluding the $q=3$ case that we know to be very different (it has just a 1RSB dynamical transition).
The numerical evaluation has been performed with $\mN=262\,144$ populations, each made of $\mM=512$ marginals.
This unbalanced choice ($\mM \ll \mN$) is due to the observation that the complexity $\Sigma(\beta,x)$ shows much stronger finite size effects in $\mN$ than in $\mM$.
The reweighing factor $r$ is \emph{dynamically} adapted during the run in order to avoid the presence of ``twins'' in a population, that would reduce the effective size of the population; the range actually spanned is $r\in[2,10]$.

In Fig.~\ref{fig:sigma_1rsb}, we plot the complexity $\Sigma(\beta=2\beta_c,x)$ for all the $q$ values studied.
We have used the same plot ranges in all panels, in order to allow a direct comparison between different $q$ values.
In each plot, we also draw the fitting quadratic curve that we use to estimate the Parisi parameter $x^*$, such that $\Sigma(x^*)=0$. The values of the estimated $x^*$ parameters are reported in Table~\ref{table}.
Errors on $\Sigma$ are large due to the fact we are measuring a very small complexity, $\Sigma \sim O(10^{-4})$.

\begin{table}[h]
\caption{1RSB parameters in the $q$-state clock model with $J_{ij}=\pm1$ couplings ($p=1/2$) on a random $3$-regular graph.}
\begin{tabular}{c|c|c|c}
	\hline
		$q$ & $x^*$ & $Q_0$ & $Q_1$\\
	\hline
		2 & 0.45(1) & 0.497(4) & 0.748(1)\\
		4 & 0.47(2) & 0.506(9) & 0.750(2)\\
		5 & 0.48(1) & 0.427(5) & 0.700(1)\\
		6 & 0.51(3) & 0.499(9) & 0.685(3)\\
		7 & 0.47(2) & 0.447(8) & 0.666(2)\\
		8 & 0.46(2) & 0.449(7) & 0.664(2)\\
	\hline
\end{tabular}
\label{table}
\end{table}

The overlaps $Q_0$ and $Q_1$ as a function of $x$ are shown in Fig.~\ref{fig:overlap_1rsb} for all the $q$ values studied.
Statistical errors on the overlaps at a given $x$ are smaller than the symbol size; thus the uncertainty reported in Tab.~\ref{table} is completely due to the error on the estimation of $x^*$.
We notice that data for $q=2$ and $q=4$ in Fig.~\ref{fig:overlap_1rsb} perfectly coincide (and this is expected for the reasons explained above). More interestingly is that also data for $q=7$ and $q=8$ coincide: this seems to suggests that, after the ``transient'' values $q=5$ and $q=6$, the $q$-state clock model converges immediately to its large $q$ limit, the XY model.

A comparison with the 1RSB solution for the corresponding fully connected model can be done only in the Ising case ($q=2$); for the SK model the 1RSB solution returns at $T=T_c/2$ the following parameters~\cite{Parisi1980a}:
\[
	x^*=0.28 \, , \quad Q_0=0.213 \, , \quad Q_1=0.619\,,
\]
which are rather different from the ones describing the 1RSB solution in the $q=2$ sparse case (see Table~\ref{table}).
So the solutions we are studying are quite far from those on the fully connected topology.
Nonetheless, similarly to what has been observed in the fully connected case \cite{NobreSherrington1986}, we notice that the 1RSB parameters in Table~\ref{table} seem to vary little for $q \ge 5$; actually they are compatible with $q$-independent values within the error bars. Only the value $q=6$ shows a peculiar behavior, maybe due to some reminiscence of the $q=3$ case. Given the very fast convergence in $q$, and willing to have a precautionary attitude, we can safely take the $q=8$ clock model as a very good approximation to the XY model, even in the RSB low-temperature phase. Also, the $d=3$ cubic lattice case in the Migdal-Kadanoff RG approximation~\cite{IlkerBerker2014} shows very similar results, with $q=2$ and $q=4$ having the same critical exponents, $q=3$ showing a peculiar behavior, and finally $q \gtrsim 5$ converging to the asymptotic values of the XY model.

\section{Conclusions}

In this work we have studied analytically the disordered XY model on random regular graphs (Bethe lattice) with different disorder distributions.
We have used the cavity method, that provides the correct answer for models with a locally tree-like topology.
Given that the (replica symmetric) solution to the XY model requires to deal with $O(N)$ probability distributions on $[0,2\pi)$, with $N$ being the system size, we have chosen to approximate the XY model with the $q$-state clock model. The computational effort to solve the latter scales as $O(q^2 N)$.

First of all we have shown that both the physical observables and the critical lines of the $q$-state clock model converge to those of the XY model very quickly in $q$: exponentially in $q$ for positive temperatures ($T>0$) and following a stretched exponential in $q$ for $T=0$. In practice we believe that the clock model with a not too large number of states can be safely used both in analytical computations and numerical simulations, especially in presence of quenched disorder and positive temperatures. Indeed, the situation where the discretization becomes more evident is the very low-temperature limit in presence of a very weak disorder. Avoiding this limit, the clock model perfectly mimics the XY model.

Secondly, by using the clock model with large $q$ as a proxy for the XY model, we have computed accurate phase diagrams in the temperature versus disorder strength plane. We have used different disorder distributions and found some common features, that are apparently universal.
Among these features, the one which is markedly different from disordered Ising models is the presence of spontaneous replica symmetry breaking at zero temperature for any disorder level: indeed, even an infinitesimal amount of disorder seems to bring the system into a spin glass phase.
This new finding suggests that in the XY model it is much easier to create many different kinds of long range order (namely thermodynamic states), probably due to the continuous nature of XY variables.

Finally, we tried to study the low-temperature spin glass phase within the ansatz with one step of replica symmetry breaking. Our original aim was to make connection to what is known in the fully connected clock model. The 1RSB analytical solution is computationally very demanding. Within the accuracy we managed to achieve, the clock model with finite connectivity studied here looks quantitatively far from its fully connected version. However, as in the fully connected case, the dependence on $q$ seems to be relevant only for very small values of $q$: as soon as $q \gtrsim 5$, the model properties resemble those of the XY model.

\begin{acknowledgments}
The authors thank Giorgio Parisi for useful discussions.
This research has been supported by the European Research Council (ERC) under the European Unions Horizon 2020 research and innovation programme (grant agreement No [694925]).
\end{acknowledgments}

\appendix

\section{Susceptibility propagation}
\label{app:susc_propag}

As stated in Sec.~\ref{subsec:validity_rs_method}, one of the most used methods to detect the RS instability is to linearize the BP equations and see when RS fixed point becomes unstable under a small perturbation. This method is better known as susceptibility propagation, since the propagation of the BP perturbations allows one to compute the susceptibility in any given sample.

\subsection{Susceptibility propagation at a positive temperature}

Let us start from the $J_{ij}=\pm 1$ case at $T>0$. We use the notation for the XY model, since the corresponding equations for the clock model are easily obtained by changing all integrals over $\theta_i$'s with discrete sums. To~help the reader, we rewrite here the BP equations~(\ref{eq:self_cons_eta}):
\begin{equation}
		\eta_{i\to j}(\theta_i)=\frac{1}{\mZ_{i\to j}}\,\prod_{k\in\partial i\setminus j}\int d\theta_k\,e^{\,\beta J_{ik}\cos{(\theta_i-\theta_k)}}\,\eta_{k\to i}(\theta_k)\,,
		\label{eq:app_self_cons_eta}
\end{equation}
with $\mZ_{i\to j}$ given by:
\begin{equation*}
		\mZ_{i\to j}=\int d\theta_i\,\prod_{k\in\partial i\setminus j}\int d\theta_k\,e^{\,\beta J_{ik}\cos{(\theta_i-\theta_k)}}\,\eta_{k\to i}(\theta_k)\,.
\end{equation*}

The most generic perturbation to a cavity marginal $\eta_{i\to j}(\theta_i)$ must be such that the perturbed marginal
\[
\eta'_{i\to j}(\theta_i)=\eta_{i\to j}(\theta_i)+\delta\eta_{i\to j}(\theta_i)
\]
remains well normalized, thus implying
\[
\int d\theta_i\,\delta\eta_{i\to j}(\theta_i)=0\,.
\]
So, in the study of the instability of the RS fixed point, one has to search for the most unstable perturbation among those satisfying the above condition.

Linearization of the BP equations~(\ref{eq:app_self_cons_eta}) leads to
\begin{widetext}
\begin{equation}
\begin{aligned}
	\delta\eta_{i\to j}(\theta_i)&=\frac{1}{\mZ_{i\to j}}\sum_{k\in\partial i\setminus j}\int d\theta_k\,e^{\,\beta J_{ik}\cos{(\theta_i-\theta_k)}}\,\delta\eta_{k\to i}(\theta_k)\prod_{k'\neq k}\int d\theta_{k'}\,e^{\,\beta J_{ik'}\cos{(\theta_i-\theta_{k'})}}\,\eta_{k'\to i}(\theta_{k'})\\
	&\quad -\frac{1}{\mZ_{i\to j}^2}\,\Biggl[\,\prod_{k\in\partial i\setminus j}\int d\theta_k\,e^{\,\beta J_{ik}\cos{(\theta_i-\theta_k)}}\,\eta_{k\to i}(\theta_k)\Biggr]\\
	&\qquad \times\int d\theta_i\,\Biggl[\,\sum_{k\in\partial i\setminus j}\int d\theta_k\,e^{\,\beta J_{ik}\cos{(\theta_i-\theta_k)}}\,\delta\eta_{k\to i}(\theta_k)\prod_{k'\neq k}\int d\theta_{k'}\,e^{\,\beta J_{ik'}\cos{(\theta_i-\theta_{k'})}}\,\eta_{k'\to i}(\theta_{k'})\Biggr]\,.
	\label{eq:app_self_cons_eta_linear_pm_J}
\end{aligned}
\end{equation}
\end{widetext}
All cavity messages that appear in this expression --- as well as normalization constant $\mZ_{i\to j}$ --- have to be evaluated on the RS BP fixed point.

Eq.~(\ref{eq:app_self_cons_eta_linear_pm_J}) can be solved on a given graph or in population dynamics if one is interested only in computing the typical behavior.
In the latter case, we need to evolve a population of pairs of functions $(\eta_i(\theta_i),\delta\eta_i(\theta_i))$: marginals $\eta_i(\theta_i)$ evolve according to BP equations~(\ref{eq:app_self_cons_eta}), while perturbations $\delta\eta_i(\theta_i)$ evolve according to Eq.~(\ref{eq:app_self_cons_eta_linear_pm_J}).
We then measure the growth rate of the perturbations via the norm
\begin{equation}
	|\delta\eta|\equiv\sum_{\{i\to j\}}\,\int d	\theta_i\,|\eta_{i\to j}(\theta_i)|\,,
\end{equation}
that evolves as $|\delta\eta| \propto \exp(\lambda\, t)$ for large times.
The critical point leading to RS instability is identified by the condition $\lambda=0$.

In the gauge glass the linearized BP equations read
\begin{widetext}
\begin{equation}
\begin{aligned}
	\delta\eta_{i\to j}(\theta_i)&=\frac{1}{\mZ_{i\to j}}\sum_{k\in\partial i\setminus j}\int d\theta_k\,e^{\,\beta J\cos{(\theta_i-\theta_k-\omega_{ik})}}\,\delta\eta_{k\to i}(\theta_k)\prod_{k'\neq k}\int d\theta_{k'}\,e^{\,\beta J\cos{(\theta_i-\theta_{k'}-\omega_{ik'})}}\,\eta_{k'\to i}(\theta_{k'})\\
	&\quad -\frac{1}{\mZ_{i\to j}^2}\,\Biggl[\,\prod_{k\in\partial i\setminus j}\int d\theta_k\,e^{\,\beta J\cos{(\theta_i-\theta_k-\omega_{ik})}}\,\eta_{k\to i}(\theta_k)\Biggr]\\
	&\qquad \times\int d\theta_i\,\Biggl[\,\sum_{k\in\partial i\setminus j}\int d\theta_k\,e^{\,\beta J\cos{(\theta_i-\theta_k-\omega_{ik})}}\,\delta\eta_{k\to i}(\theta_k)\prod_{k'\neq k}\int d\theta_{k'}\,e^{\,\beta J\cos{(\theta_i-\theta_{k'}-\omega_{ik'})}}\,\eta_{k'\to i}(\theta_{k'})\Biggr]\,.
	\label{eq:app_self_cons_eta_linear_gg}
\end{aligned}
\end{equation}
\end{widetext}

\subsection{Susceptibility propagation at zero temperature}

At zero temperature the situation is more complicated and important differences arise between the $q$-state clock model with discrete variables and the XY model with continuous variables.

Let us recall the zero temperature version of the BP equations for the $J_{ij}=\pm 1$ XY model:
\begin{equation}
	h_{i\to j}(\theta_i)\cong\sum_{k\in\partial i\setminus j}\max_{\theta_k}{\left[h_{k\to i}(\theta_k)+J_{ik}\cos{(\theta_i-\theta_k)}\right]}
	\label{eq:app_self_cons_h_T_0_pm_J}
\end{equation}
where $h_{i\to j}(\theta_i)$ is defined up to an additive constant such that $\max_{\theta_i}[h_{i\to j}(\theta_i)]=0$, as we discuss in Sec.~\ref{sec:bp_cavity_method_rs}.
Linearizing Eq.~(\ref{eq:app_self_cons_h_T_0_pm_J}) we get the following equation for the evolution of perturbations
\begin{equation}
	\delta h_{i\to j}(\theta_i)\cong\sum_{k\in\partial i\setminus j}\delta h_{k\to i}(\theta_k^*(\theta_i))\,,
	\label{eq:app_self_cons_h_T_0_linear_pm_J}
\end{equation}
where $\theta_k^*(\theta_i)$ is given by:
\begin{equation}
	\theta_k^*(\theta_i)=\operatorname{arg}\,\max_{\theta_k}{\left[h_{k\to i}(\theta_k)+J_{ik}\cos{(\theta_i-\theta_k)}\right]}\,.
	\label{eq:theta_star}
\end{equation}

In practice we never solve the equations for the XY model; we always solve those for the $q$-state clock model, where the only difference is that the maximum in Eqs.~(\ref{eq:app_self_cons_h_T_0_pm_J}) and~(\ref{eq:theta_star}) must be taken over the discrete set of $q$ possible values for $\theta_k$.

A natural question is whether the solution to these equations changes smoothly with $q$ in the limit of very large $q$.
We find that the marginals in the $q$-state clock model with $q$ large are very close to those in the corresponding XY model: thus Eq.~(\ref{eq:app_self_cons_h_T_0_pm_J}) for the marginals can be used safely, and the XY model well approximated by the clock model with moderately large values for $q$.

On the contrary, perturbations in the clock model at $T=0$ evolve in a completely different way with respect to what happens in the XY model.
Indeed, due to the fact the maximum in Eq.~(\ref{eq:theta_star}) is taken over a discrete set, it can happen that perturbations obtained from Eq.~(\ref{eq:app_self_cons_h_T_0_linear_pm_J}) have the same value for any $\theta_i$. And this in turn corresponds to a null perturbation.

To understand the last statement, one has to remember that the correct normalization for each cavity marginal is enforced by the condition that the maximum of $h_{i\to j}(\theta_i)$ is zero, and this must be true also for the perturbed marginal. This implies that the perturbation must be zero in $\theta_i^\text{max}=\text{argmax}_{\theta_i} [h_{i\to j}(\theta_i)]$.
We enforce this condition by shifting the perturbations obtained from Eq.~(\ref{eq:app_self_cons_h_T_0_linear_pm_J}) as follows
\[
\delta h_{i\to j}(\theta_i) \;\leftarrow\; \delta h_{i\to j}(\theta_i) - \delta h_{i\to j}(\theta_i^\text{max})\,.
\]
Consequently a constant perturbation generated by Eq.~(\ref{eq:app_self_cons_h_T_0_linear_pm_J}) becomes a null perturbation after the shift.

So, in the $T=0$ clock model, the perturbations evolve exactly as in the Ising model~\cite{MoroneEtAl2014a}, where perturbations divide in two groups: null perturbations and nonnull perturbations. Since the norm of the nonnull perturbations changes a little and remains finite during the evolution, the instability of the RS BP fixed point is mainly determined by the evolution of the \emph{fraction} of nonnull perturbations.

Nonetheless, we have observed that such a fraction shrinks with a rate that depends on $q$, and in the $q\to\infty$ limit this fraction seems to remain finite.
So, we conclude that the shrinking of the fraction of nonnull perturbations is a direct consequence of the discretization, and it is not present in the XY model.

The fact that perturbations evolve in a drastically different way in the XY model and in the clock model for any value of $q$, may suggest the latter model is unfit to describe the fluctuations of the XY model, and thus its physical behavior in the $T=0$ limit.
Luckily enough there is a way out to this problem.

The solution is to use the BP equations~(\ref{eq:app_self_cons_h_T_0_pm_J}) in a fully discretized way, but computing the maximum in Eq.~(\ref{eq:theta_star}) on the reals.
Given that the argument of the argmax in Eq.~(\ref{eq:theta_star}) is defined only on $q$ discrete points, we interpolate with a parabola the three points around the maximum, i.\,e. the point where the argument achieves its maximum and the two nearby.
A similar interpolation is performed on the function $\delta h_{k\to i}(\cdot)$ that needs to be evaluated at the value $\theta_k^*(\theta_i)$ that no longer belongs to the discrete set.
By proceeding this way, we obtain perturbations that shrink, but never become exactly zero.
Thus the critical point can be computed just by checking the evolution of the norm of the perturbations as in the $T>0$ case.

\bibliographystyle{apsrev4-1}
\bibliography{myBiblio}

\end{document}